\newif\ifdraft\drafttrue
\newif\ifcolor\colortrue
\newif\iftechreport\techreporttrue

\iftechreport
\documentclass[onecolumn]{article}
\pdfoutput=1
\usepackage{epsfig}
\usepackage{amssymb}
\usepackage{amsmath}
\usepackage{amsfonts}
\usepackage[margin=1.0in]{geometry}
\else
\documentclass[letter]{sig-alternate-10pt}
\fi

\usepackage[table]{xcolor}
\definecolor{pennblue}{cmyk}{1,.65,0,.3}
\definecolor{pennred}{cmyk}{0,1,.65,.34}

\definecolor{williamspurple}{RGB}{89,23,128}
\definecolor{williamsgold}{RGB}{255,213,0}

\usepackage{arydshln}
\usepackage{booktabs}
\usepackage{cmtt}

\usepackage{paralist}
\usepackage{lastpage}
\usepackage{times}
\usepackage{tikz}
\usepackage{alltt}
\usepackage{graphicx}
\usepackage{balance}
\usepackage{amsmath}
\usepackage{caption}
\usepackage{subcaption}
\usepackage{algorithm,algpseudocode}
\usepackage{enumitem}

\usepackage{verbatim}

\usepackage{tikz}
\pgfdeclarelayer{background}
\pgfdeclarelayer{foreground}
\pgfsetlayers{foreground,background,main}

\iftechreport
\else
\fi
\usepackage[export]{adjustbox}
\graphicspath{{./figures/}}

\begin{document}

\iftechreport
\title{Scalable Traffic Engineering for Higher Throughput in Heavily-loaded Software Defined Networks}
\else
\title{XXX}
\fi

\iftechreport
\author{ Che Zhang \\ SUSTech
\and Shiwei Zhang \\ SUSTech
\and Yi Wang \\ SUSTech
\and Weichao Li \\ SUSTech
\and Bo Jin \\ SUSTech
\and Ricky K. P. Mok \\ CADIA/UC San Diego
\and Qing Li \\ SUSTech
\and Hong Xu \\ CityU of Hong Kong
}

\else
\numberofauthors{1}
\author{Paper \#XXX, \pageref{ConcPage} pages}

\fi

\iftechreport
\date{}
\else
\date{}
\fi

\maketitle

\begin{abstract}
Existing traffic engineering (TE) solutions performs well for software defined network (SDN) in average cases. However, during peak hours, bursty traffic spikes are challenging to handle, because it is difficult to react in time and guarantee high performance even after failures with limited flow entries.

Instead of leaving some capacity empty to guarantee no congestion happens due to traffic rerouting after failures or path updating after demand or topology changes, we decide to make full use of the network capacity to satisfy the demands for heavily-loaded peak hours. The TE system also needs to react to failures quickly and utilize the priority queue to guarantee the transmission of loss and delay sensitive traffic. We propose TED, a scalable TE system that can guarantee high throughput in peak hours. TED can quickly compute a group of maximum number of edge-disjoint paths for each ingress-egress switch pair. We design two methods to select paths under the flow entry limit. We then input the selected paths to our TE to minimize the maximum link utilization. In case of large traffic matrix making the maximum link utilization larger than 1, we input the utilization and the traffic matrix to the optimization of maximizing overall throughput under a new constrain. Thus we obtain a realistic traffic matrix, which has the maximum overall throughput and guarantees no traffic starvation for each  switch pair. Experiments show that TED has much better performance for heavily-loaded SDN and has $10\%$ higher probability to satisfy all ($> 99.99\%$) the traffic after a single link failure for G-Scale topology than Smore under the same flow entry limit.
\end{abstract}

\newcommand{\jnfpar}[1]{\ \\[-.75em] \textit{#1.}~}

\section{Introduction}
\label{sec:introduction}

CAIDA~\cite{ref:inferring} observed that congestion was not widespreaded on the peer/provider interdomain links during their measurement period.
However, Akamai said that on Dec. 11, 2018, the volume of data passed 72 Tbps, which equated to delivering more than 10 million DVDs per hour, was
largely attributed to streaming of live sports events, etc \cite{akamai}.

Several solutions work well in non-peak hours. However, they are not suitable for peak hours to handle high volume traffic bursts and guarantee high performance even after failures. We can classify these solutions into three categories. The first category is to proactively consider failures when formulating the TE problem. However, these solutions neither may not be scalable to large network due to exponential number of all possible failure scenarios~\cite{ref:joint,ref:teavar} nor require to reserve significant portion of network capacity in order to guarantee no congestion happens for arbitrary k faults with rescaling~\cite{ref:ffc}. The second way is to periodically update the paths and weights according to current traffic demand matrix, but it also needs to reserve bandwidth to prevent network congestion or network looping caused by path update events~\cite{ref:swan}. The third type is to pre-calculate and deploy a large number of static paths to avoid a series of problems caused by path updates, e.g., Smore's semi-oblivious routing~\cite{ref:smore}. They often use minimizing the maximum link utilization (MLU) as the optimization target. Although minimizing the maximum link utilization can make network load balanced, it will limit the whole throughput and lead to large amount of packet loss when the load is heavy as it requires a constrain to satisfy all the demands. Maximizing the whole throughput is a natural objective for heavy load, but optimize the whole throughput only may lead to traffic starvation for some ingress-egress switch pairs.

Moreover, Google found that off-the-shelf switch chips impose hard limits on the size of each flow table~\cite{ref:b4andafter}, which also limit
the number of deployable static paths especially for large scale networks as traffic engineering systems usually implement flow group matching using ACL tables to leverage their generic wildcard matching capability. Motivated by~\cite{ref:b4andafter} and the design of the third type of TE method~\cite{ref:smore}, we design TED to solve the above problems. TED utilizes the advantages of both optimizations to maximizing the throughput with no starvation for each switch pair guaranteed.

TED is a TE system that is fast, scalable, and simple to be used. TED typically includes three phases: I) path set computation, II) path selection, and III) weight computation \& bandwidth allocation. By extending Dinic's algorithm \cite{ref:dinic}, TED can compute a group of maximum number of edge-disjoint paths between each ingress-egress switch pair with time complexity $O(n^2m)$ (The algorithm is called ``Custom" to remind users that they can also use other path computation methods in our TED system). Using the limitation of the number of flow entries for each switch (router) in the WAN network topology, TED can search the maximum path budget with time complexity $O(n^2)$ (``hardnop"). Furthermore, we design a two-step path selection method (``program") to select paths using 0-1 integer linear programming (Fig. 10c).

For TE optimization, the goal of TED is to minimize the maximum link utilization. We do not directly apply the weights when its result $Z>1$ ($Z$ stands for the maximum link utilization) to avoid network congestion. Instead, Phase IV of TED is triggered. Input $TM$ and $Z$ to a new TE optimization with objective of maximizing the overall throughput under link capacity constraint, weights and bandwidth allocation are re-computed under new constraint $TM/Z \leq TM' \leq TM$. Phase IV not only guarantees no congestion even when $Z>1$, but also can fully utilize the network capacity under the guarantee that each switch pair can at least meet $1/Z$ of its demand. Another feature of TED is customizable. The operator or network slicing user can apply other path set computation algorithms, limit the use of flow entries,
or reserve a portion of link capacity for robustness.

Evaluation shows TED has much better performance for heavily-loaded networks (Fig. 8a) and has 10\% more possibility to satisfy all the traffic ($P(T>99.99\%)$) after single link failures than Smore (Fig. 11a, under the same flow entry limit and topology). TED has only less than 10\% $s-t$ pairs assigned an average path length, which is longer than the maximum average path length computed by Smore (Fig.11b), but Smore uses 33\% or more flow entries in some switches (Fig. 7a, 7b).

\section{Background}
\label{sec:background}

The most commonly used way to solve the multi-commodity flow problem \cite{ref:networkflow}, liner programming takes the input of $(s_i,t_i,d_i)$ pairs and the paths between $(s_i,t_i)$, where $s_i$ and $t_i$ are the $i^{th}$ pair of ingress ($s$) and egress ($t$) routers with demand $d_i$ (refer to Fig. \ref{fig:sys} Phase II).

\begin{figure}[!htbp]
\centering
\includegraphics[width=0.6\linewidth]{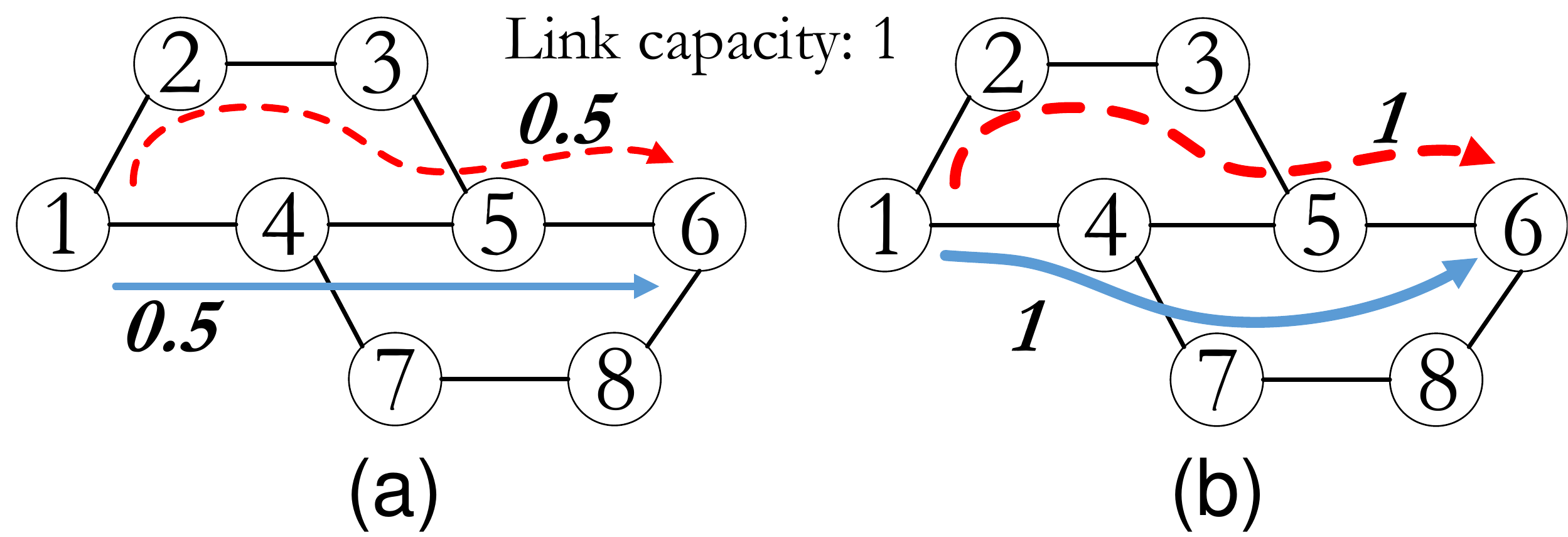}
\caption[]{Comparison of bandwidth allocation between selecting two shortest paths (a) and two edge disjoint paths (b).}
\label{fig:moti}
\end{figure}

\begin{figure}[!htbp]
\hspace*{-0.5cm}\includegraphics[width=1.1\linewidth]{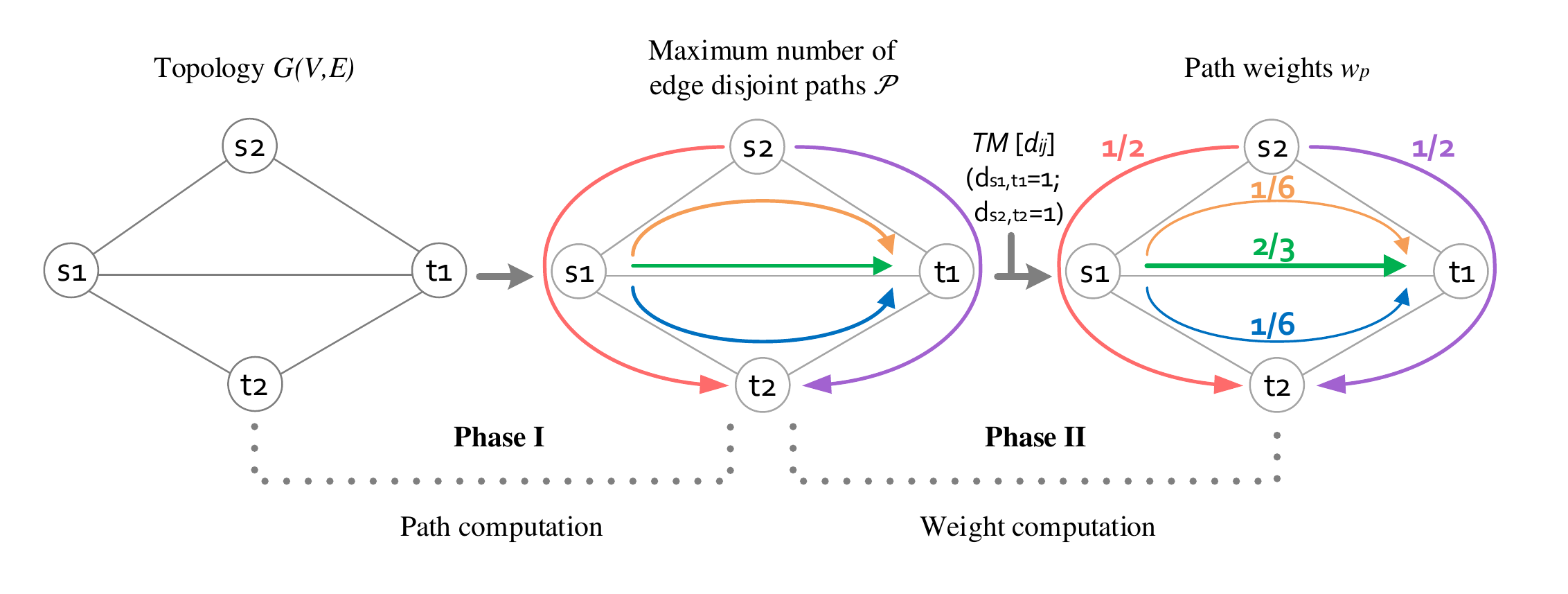}
\caption[]{Using edge-disjoint paths computation method in previous TE system architecture.}
\label{fig:sys}
\end{figure}

We use a simple example to illustrate the influence of selection of paths to bandwidth allocation result of TE and reliability of the network.
Fig. \ref{fig:moti} (a) shows that if we select two shortest paths, indicated with red dotted line ($1 \rightarrow 2 \rightarrow 3 \rightarrow 5 \rightarrow 6$) and blue solid line ($1 \rightarrow 4 \rightarrow 5 \rightarrow 6$ ) for $s-t$ pair $(1,6)$. Because the two paths share the link $5 \rightarrow 6$, each of them can only obtain half of link capacity. SWAN \cite{ref:swan} finds that each ingress-egress switch pair needs to have at least 15 shortest paths to fully utilize the overall network throughput. In contrast, by selecting two edge-disjoint paths as shown in Fig \ref{fig:moti} (b), the available bandwidth of both paths is doubled. Even though the two disjoint paths are not the shortest ones, they are more robust to single link failures.

We further analyzed Google G-Scale Topology \cite{ref:b4} to verify our previous conclusion. We selected $k$-shortest paths, where $k$ is the maximum number of edge-disjoint paths, for any $s-t$ pair. We found that, on average, around five $s-t$ pairs had at least two paths that are vulnerable to any single link failure. In the worst case, 13.6\% pairs had at least two paths with shared bottleneck, which is undesirable for WANs
with high link utilization. Although 26.3\% of single link failures did not impact on more than one $k$-shortest path for any node pair, the remaining 73.7\% single link failure can cause significant impact to the network because of the shared link among selected paths.
A suitable path selection is the key of ensuring high reliability and availability in WANs. Specifically, selecting edge-disjoint paths can make network more fault-tolerance and even improve the overall network throughput with a chance of a slight sacrifice of latency.

\begin{figure}[!htbp]
\includegraphics[width=1.0\linewidth]{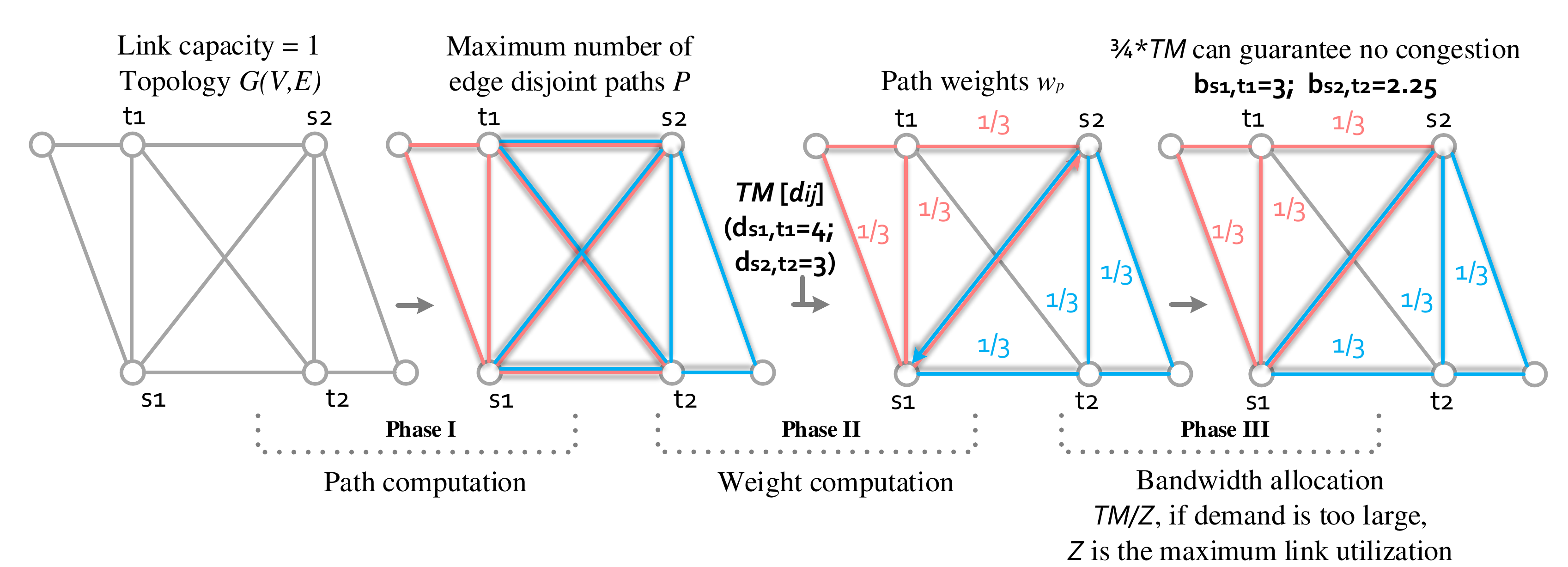}
\caption[]{When $TM$ is large, using the weights computed by minimize the maximum link utilization may lead to congestion. Naive way to limit the bandwidth of each path as $TM/Z$ may lead to inefficiency using of network capacity.}
\label{fig:sysnew1}
\end{figure}

Next, we will show why we cannot directly apply traditional TE system. Fig. \ref{fig:sys} illustrate our attempt to combine edge disjoint paths computation with traditional two-phase TE system model. In phase I, we select the maximum number of edge disjoint paths. After that, we employ the paths in phase I and using minimizing maximum link utilization optimization to compute the bandwidth allocation and corresponding weights in phase II. It seems to work well, but first, WANs need to guarantee all-to-all connections and when the number of flow entries are limited, we cannot simply select commonly used 3 or 4 paths \cite{ref:joint,ref:b4,ref:smore} between each ingress-egress switch pair. It is because the result can be either limited flow entries cannot support those paths especially in large WANs or path diversity decreases due to the ``3 or 4" limitation.

With development of network measurement like \cite{ref:bwe} and SDN-based system used in Internet like Google's  Espresso \cite{espresso} and Facebook's Edge fabric \cite{edgefabric}, we believe that the estimated or measured $TM$ can be more accurate. TED intends to use minimizing the \emph{maximum link utilization} ($Z$) optimization to compute bandwidth allocation and corresponding weights, as it has a constraint to meet all demands in $TM$ and balances the traffic. Furthermore, the implementation of such optimization is easier than minimizing the overall network congestion which is a convex optimization and hard to select a proper penalty function to use \cite{ref:joint}.

The $TM$ increases as the use and expectation of high quality (High definition, 4K, or even 8K) video streaming, IPTV, and video conferencing \cite{edgefabric}. Then $Z$ may be larger than 1 as the optimization of minimizing the maximum link utilization has no link capacity constraint (in this case, the large $TM$ may not be satisfied and the optimization will have no result that satisfies all constraints). Although when $Z>1$ we can simply allocate bandwidth as $TM/Z$ without changing the weights to guarantee no congestion happens, the network may not be fully used as shown in the example of Fig. \ref{fig:sysnew1}.

\section{Design Overview}
\label{sec:design}

\begin{figure*}[!ht]
\centering
\includegraphics[width=1.0\textwidth]{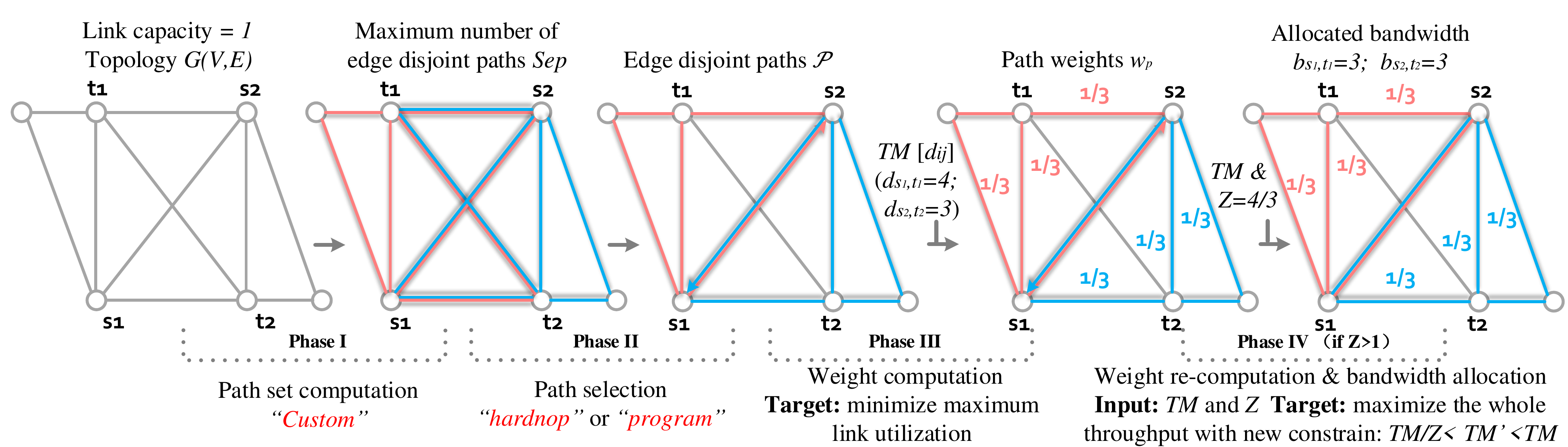}
\caption[]{TED architecture.}
\label{fig:sysnew4}
\end{figure*}

In this section, we present the design of TED. Fig. \ref{fig:sysnew4} depicts the architecture of TED in a four-phase TE system model.

We compute the maximum number of edge disjoint paths between each ingress-egress router pair as the path set in Phase I. We add a path selection phase II to deal with flow entries' limitation and weight re-computation \& bandwidth re-allocation phase IV to deal with large $TM$ that makes $Z>1$.

Our prioritised objective is to 1. meet the $TM$, 2. minimize failure impact, 3. guarantee at least $TM/Z$ is satisfied, when $TM$ can not be satisfied without causing congestion and 4. maximize overall network throughput. In other words, minimizing  network congestion so that the impacted traffic are as little as possible after failures. And such traffic can still be satisfied almost maximally after rerouting or using the failover mechanism. Even for network failures in peak hours, we can still guarantee the transmission of loss and delay sensitive traffic utilizing priority queue and fast failover as most video traffic are loss tolerant. Moreover, TED can recompute the paths, weights and allocated bandwidth quickly to further recover the whole network maximally.

The detailed explanation of each phase are as follows.

\subsection{Compute A Group of Maximum Number of s-t Edge-Disjoint Paths}

Select paths for demands: $(s_i,t_i,d_i), 1 \leq i \leq n*(n-1)$.
Finding all groups of $s-t$ edge-disjoint paths is too complicated and requires exponential time. Therefore, we look for a group of edge-disjoint paths with maximum total number.

A lot of fast algorithms have been developed to solve the maximum flow problem for unit capacity, undirected networks \cite{ref:dinicmit}.
Papers by Karzanov \cite{ref:karza} and Even and Tarjan \cite{ref:flowbook} showed that, for unit capacity networks (directed or undirected), a method called blocking flows invented by Dinic \cite{ref:dinic}, solves the maximum flow problem in $O(m$ min $({n^{2/3}, m^{1/2}, v}))$ time (v is the value of the maximum flow).

We extend Dinic's algorithm \footnote{Dinic's algorithm is a strongly polynomial algorithm for computing the maximum flow in a flow network. The algorithm runs in $O(n^2m)$ time and each augmenting path used in the algorithm is the shortest one.} \cite{ref:dinic} to compute a group of maximum number of edge disjoint paths for our undirected graph with unit capacity (the weight of each edge is 1).

\subsubsection{Problem description}

Given an undirected graph and two nodes (i.e., the source $s$ and the destination $t$) in it, the problem is to find out the maximum number of edge-disjoint paths from the source to the destination. Two paths are said edge-disjoint if they don't share any edge.

This problem can be solved by reducing it to maximum flow problem. Following are steps:
\begin{enumerate}[label=(\roman*)]
\item Transform the original undirected graph into a symmetric directed graph;
\item Assign unit capacity to each edge;
\item Consider the given source and destination as source and sink in a flow network;
\item Run classic maximum flow algorithms to find the maximum flow from source to sink;
\item The maximum flow is equal to the maximum number of edge-disjoint paths.
\end{enumerate}

\subsubsection{Path construction}

Now we get the final flow matrix $f^*$ from the maximum flow algorithm. Note that multiple optimal solutions may exist, but the maximum flow value is the same.

With $f^*$, we are next to construct edge-disjoint paths. Note that even for a single flow matrix $f^*$, different path sets may be constructed.

The process of the path construction is quite simple:
\begin{enumerate}[label=(\roman*)]
\item Starting from the source, for each node $u$, find the next \emph{unvisited} edge $(u,v)$ such that $f(u,v)=1$, and then move forward to $v$. Once the sink is reached, one path is constructed.
\item Repeat the previous process until all edges are visited.
\end{enumerate}

\begin{figure}
\centering
\includegraphics[width=0.6\linewidth]{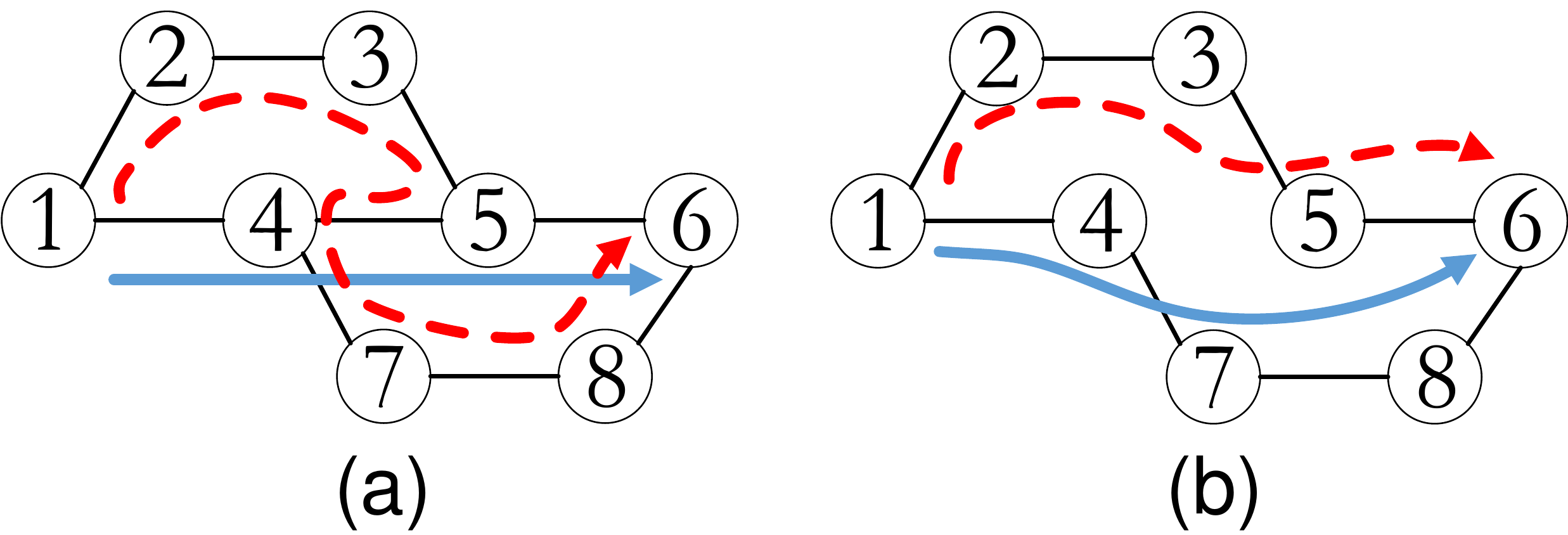}
\caption[]{The augmenting paths used in Dinic's algorithm to compute the max flow of $1 \rightarrow 6$ for original graph (a) and the max flow network of $1 \rightarrow 6$ (b) in which the edge $(4,5)$ (whose two directions are both used) is counteracted.}
\label{fig:alg}
\end{figure}

Note that we can not use the augmenting paths directly, because sometimes, two augmenting paths may pass  through the positive and negative direction of the same undirected edge as shown in Fig. \ref{fig:alg} (a). For the max flow network which has the skew symmetry property is shown in Fig. \ref{fig:alg} (b).

When $disp_i=2$, we can use Suurballe algorithm \cite{ref:suurballe} to find the shortest pairs of edge disjoint paths which has minimum total length. Notice that our path computation method can not guarantee the minimum total length.

\subsection{Select Paths}

As the limitation of switch memory (e.g., TCAM) and the fairness among each $s-t$ pair, we select paths based on path budget \cite{ref:smore}. First, we sort the paths based on their weights. For our maximum number of edge disjoint paths, we use the path length as the weight in order to reduce the latency. Second, different from previous work \cite{ref:smore}, we get the maximum path budget $K$ under flow entries' limitation with time complexity $O(n^2)$. Third, we select the first $min(K,X)$ paths for each $s-t$ pair ($X$ stands for the maximum number of edge disjoint $s-t$ paths).

For the second step, considering normally $K$ is small, we can check from $i=1$ path for each $s-t$ pair and record the used number of flow entries of each switch to a vector. Judge whether they are under the flow entries' limitation. If no, stop and $K=i-1$, otherwise, $i++$, go to next cycle. To be faster, in each cycle, we can reuse the previously recorded vector and only need to add the used flow entries of $i^{th}$ path for each $s-t$ pair to that vector.

We also design a two-step path selection method (``program") to make full use of limited flow entries. Step 1. find the result $F$ of maximize minimum $nop_{s,t}$ (means number of paths between $s$ and $t$) shown in Opt. \eqref{opt:maxminnopst}.

\begin{table}[t]\small
\caption[]{Summary of notation}
\label{table:symbols}
\begin{center}
\begin{tabular}{l l}
\hline
Variable & Description \\
\hline
$\mathbb{G}(\mathbb{V},\mathbb{E})$ & network with vertices $\mathbb{V}$ and directed edges $\mathbb{E}$\\
$c_{e_k}$ & capacity of $k^{\text{th}}$ edge, $e_k \in \mathbb{E}$\\
$\mathbb{P}$ & a set of selected edge disjoint paths\\
$\mathbb{TM}$ & bandwidth demand matrix\\
$h_{v_o}$ & limitation of number of flow entries in $o^{\text{th}}$ switch \\
\hline
$\mathbb{D}$ & bandwidth demand set, $\forall (s,t) \in \mathbb{D}, \mathbb{TM}[s,t] \neq 0$\\
$(s_i,t_i)$ & $i^{\text{th}}$ ingress-egress switch pair in $\mathbb{D}$\\%
$\mathbb{P}(s_i,t_i)$ & edge disjoint paths between $s_i$ and $t_i$\\
$L[p_j,e_k]$ & if
path $p_j$ passed through $e_k$, $L[p_j,e_k]=1$, else 0\\
$R[p_j,v_o]$ & if
path $p_j$ passed through $v_o$, $R[p_j,v_o]=1$, else 0\\
$d_{s_i,t_i}$ & bandwidth demand of $(s_i,t_i)$, $d_{s_i,t_i}=\mathbb{D}(s_i,t_i)$\\
$l_{e_k}$ & overall used bandwidth of link $e_k$\\
$q_{v_o}$ & used flow entries of $o^{\text{th}}$ switch\\
\hline
$a_{s_i,t_i}^{p_j}$ & 1 for selecting
$j^{\text{th}}$ path of $(s_i,t_i)$, else 0\\
$b_{s_i,t_i}^{p_j}$  & allocated bandwidth for path $p_j$ of $(s_i,t_i)$ \\
$w_{s_i,t_i}^{p_j}$  & weight for path $p_j$ of $(s_i,t_i)$\\
\hline
\end{tabular}
\end{center}
\end{table}
\begin{equation*}
  \tag{1}\label{opt:maxminnopst}
\begin{aligned}
&\text{max} & & min_{(s_i,t_i) \in \mathbb{D}}{(nop_{s_i,t_i})} \\
& \text{s.t.} & & a_{s_i,t_i}^{p_j} \in \{0,1\}\\
&&& q_{v_o}=\sum_{(s_i,t_i) \in \mathbb{D}}{\sum_{p_j \in \mathbb{P}(s_i,t_i)}{a_{s_i,t_i}^{p_j}*R[p_j,v_o]}} \leq h_{v_o}, \forall v_o \in \mathbb{V}\\
&&& nop_{s_i,t_i}=\sum_{p_j \in \mathbb{P}(s_i,t_i)}{a_{s_i,t_i}^{p_j}}, \forall (s_i,t_i) \in \mathbb{D}\\
\end{aligned}
\end{equation*}
\begin{equation*}
  \tag{2}\label{opt:maxsumnopst}
\begin{aligned}
&\text{max} & & \sum_{(s_i,t_i) \in \mathbb{D}}{nop_{s_i,t_i}} \\
& \text{s.t.} & & a_{s_i,t_i}^{p_j} \in \{0,1\}\\
&&& q_{v_o}=\sum_{(s_i,t_i) \in \mathbb{D}}{\sum_{p_j \in \mathbb{P}(s_i,t_i)}{a_{s_i,t_i}^{p_j}*R[p_j,v_o]}} \leq h_{v_o}, \forall v_o \in \mathbb{V}\\
&&& F \leq nop_{s_i,t_i}=\sum_{p_j \in \mathbb{P}(s_i,t_i)}{a_{s_i,t_i}^{p_j}}, \forall (s_i,t_i) \in
\mathbb{D}
\end{aligned}
\end{equation*}
Step 2. input $F$ to find the path selection result of maximize sum of $nop_{s,t}$ with a new constrain $F \leq nop_{s,t}$ to guarantee each $s-t$ pair at least has $F$ paths (Opt. \eqref{opt:maxsumnopst}). The symbols and their explanations are shown in Table \ref{table:symbols}.

\subsection{TE Optimization Algorithm}

We use linear programming to compute the bandwidth allocation result of TE. Its input includes the edge disjoint paths $\mathbb{P}$ computed by the above path selection subsystem, the bandwidth demand matrix $\mathbb{TM}$ and the network topology $\mathbb{G}(\mathbb{V},\mathbb{E})$. The output
is allocated bandwidth $b_{s_i,t_i}^{p_j}$ for each path of each $s_i-t_i$ pair. We can obtain the weight $w_{s_i,t_i}^{p_j}$ after normalization (
$w_{s_i,t_i}^{p_j}=\frac{b_{s_i,t_i}^{p_j}}{b_{s_i,t_i}},\sum_{p_j \in \mathbb{P}(s_i,t_i)}{w_{s_i,t_i}^{p_j}}=1$).

We use minimize max($l_e/c_e$) (Opt. \eqref{opt:minmaxlu}) as the target of Phase III. If the bandwidth allocation result makes some $l_e>c_e$ (set $Z=max(l_e/c_e)$), Phase IV is triggered and we input $TM$ and $Z$ to the TE with target of maximize the whole throughput and the bandwidth constrain will be changed to $d_{s_i,t_i}/Z \leq b_{s_i,t_i} \leq d_{s_i,t_i}$ (Opt. \eqref{opt:maxT}). Thus the bandwidth allocation result will also hold the property: max($l_e$)=1.

\begin{equation*}
  \tag{3}\label{opt:minmaxlu}
\begin{aligned}
&\text{min} & & max_{e_k \in \mathbb{E}}{(l_{e_k}/c_{e_k})} \\
& \text{s.t.} & & 0 \leq b_{s_i,t_i}^{p_j},  \; \forall (s_i,t_i) \in \mathbb{D}, p_j \in \mathbb{P}(s_i,t_i)\\
&&& b_{s_i,t_i}=\sum_{p_j \in \mathbb{P}(s_i,t_i)}{b_{s_i,t_i}^{p_j}}=d_{s_i,t_i}, \forall (s_i,t_i) \in \mathbb{D}\\
&&& l_{e_k}=\sum_{(s_i,t_i) \in \mathbb{D}}{\sum_{p_j \in \mathbb{P}(s_i,t_i)}{b_{s_i,t_i}^{p_j}*L[p_j,e_k]}}, \forall e_k \in \mathbb{E}\\
\end{aligned}
\end{equation*}

\begin{equation*}
  \tag{4}\label{opt:maxT}
\begin{aligned}
&\text{max} & & \sum_{(s_i,t_i) \in \mathbb{D}}{b_{s_i,t_i}} \\
& \text{s.t.} & & 0 \leq b_{s_i,t_i}^{p_j},  \; \forall (s_i,t_i) \in \mathbb{D}, p_j \in \mathbb{P}(s_i,t_i)\\
&&& b_{s_i,t_i}=\sum_{p_j \in \mathbb{P}(s_i,t_i)}{b_{s_i,t_i}^{p_j}} \leq  d_{s_i,t_i}, \forall (s_i,t_i) \in \mathbb{D}\\
&&& d_{s_i,t_i}/Z \leq b_{s_i,t_i}, \forall (s_i,t_i) \in \mathbb{D}\\
&&& l_{e_k}=\sum_{(s_i,t_i) \in \mathbb{D}}{\sum_{p_j \in \mathbb{P}(s_i,t_i)}{b_{s_i,t_i}^{p_j}*L[p_j,e_k]}} \leq c_{e_k}, \forall e_k \in \mathbb{E}\\
\end{aligned}
\end{equation*}

\section{Evaluation}
\label{sec:evaluation}
{}

\begin{figure}[!htbp]
  \centering
\includegraphics[width=0.6\linewidth]{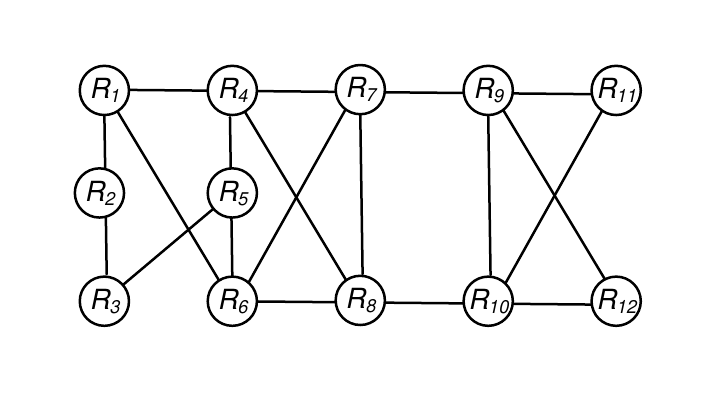}
\caption[]{Google G-Scale topology.}
\label{fig:gscale}
\end{figure}

\begin{figure*}[!htbp]
        \centering
        \begin{subfigure}[b]{0.325\textwidth}
            \centering
            \includegraphics[width=\linewidth]{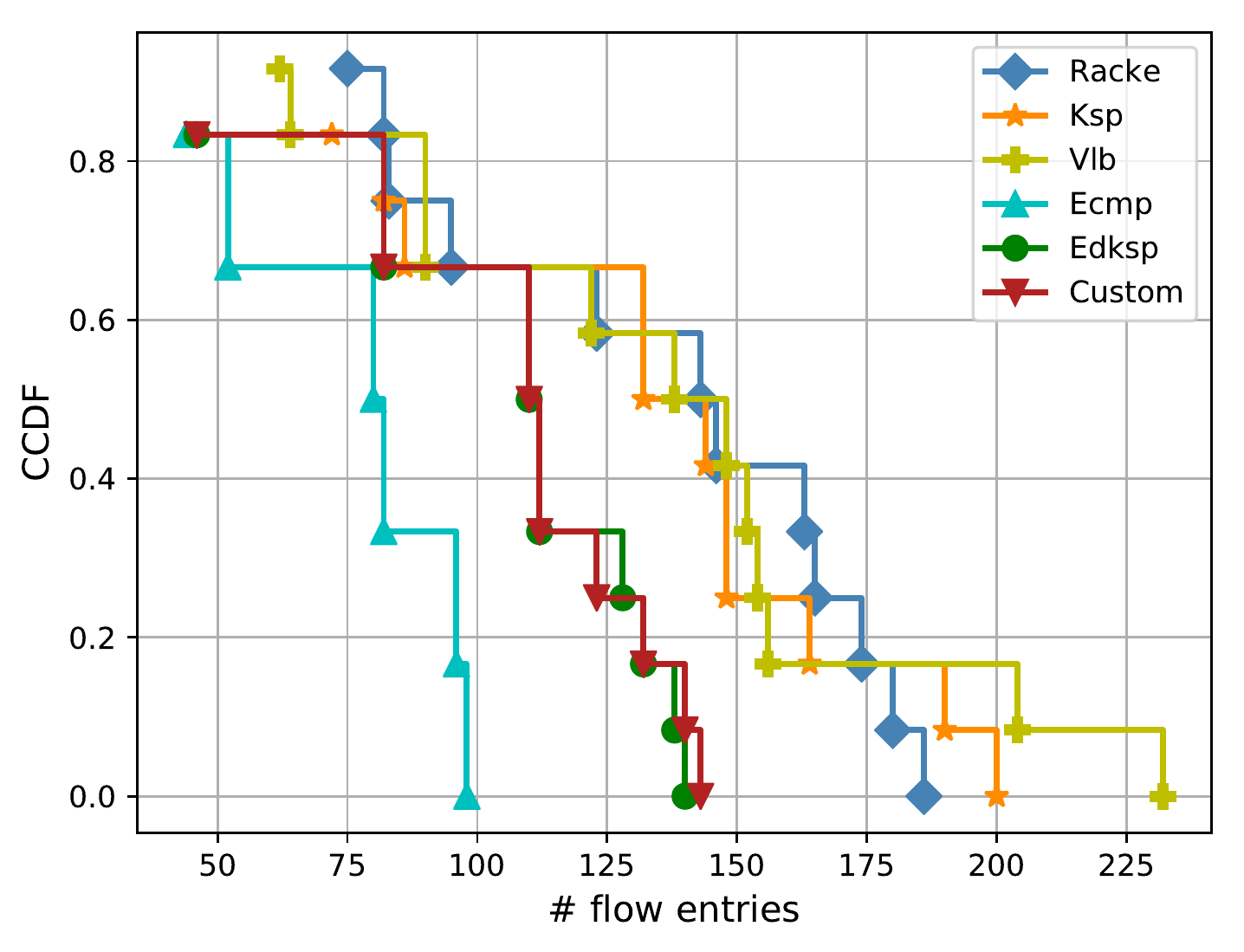}
            \caption[Network2]
            {{\small hardnop (G-Scale) }}
            \label{fig:flow-gscale-1-hardnop-ccdf}
        \end{subfigure}
        \hfill
        \begin{subfigure}[b]{0.325\textwidth}
            \centering
            \includegraphics[width=\linewidth]{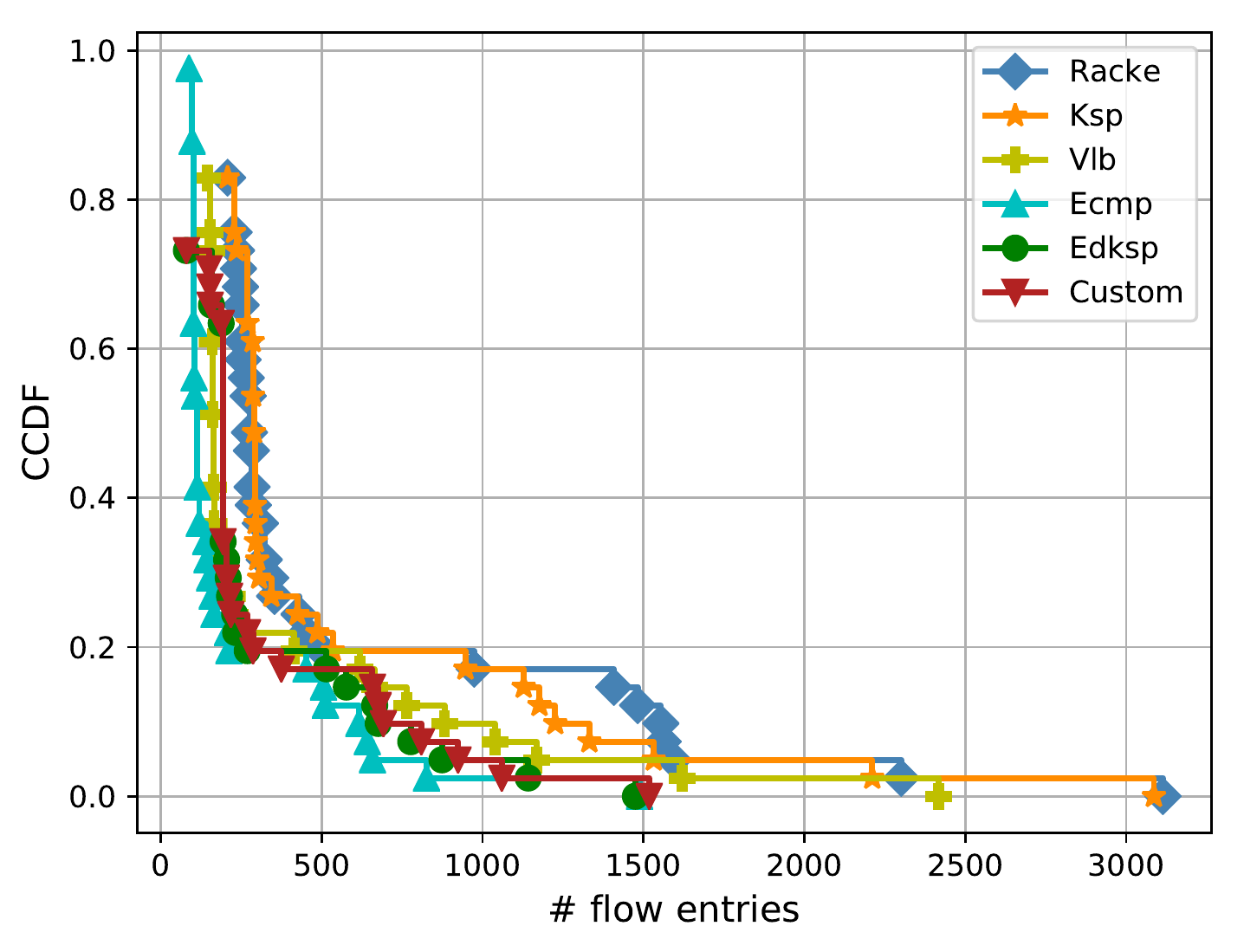}
            \caption[]
            {{\small hardnop (Cernet)}}
            \label{fig:flow-cernet-1-hardnop-ccdf}
        \end{subfigure}
        \hfill
        \begin{subfigure}[b]{0.325\textwidth}
            \centering
            \includegraphics[width=\linewidth]{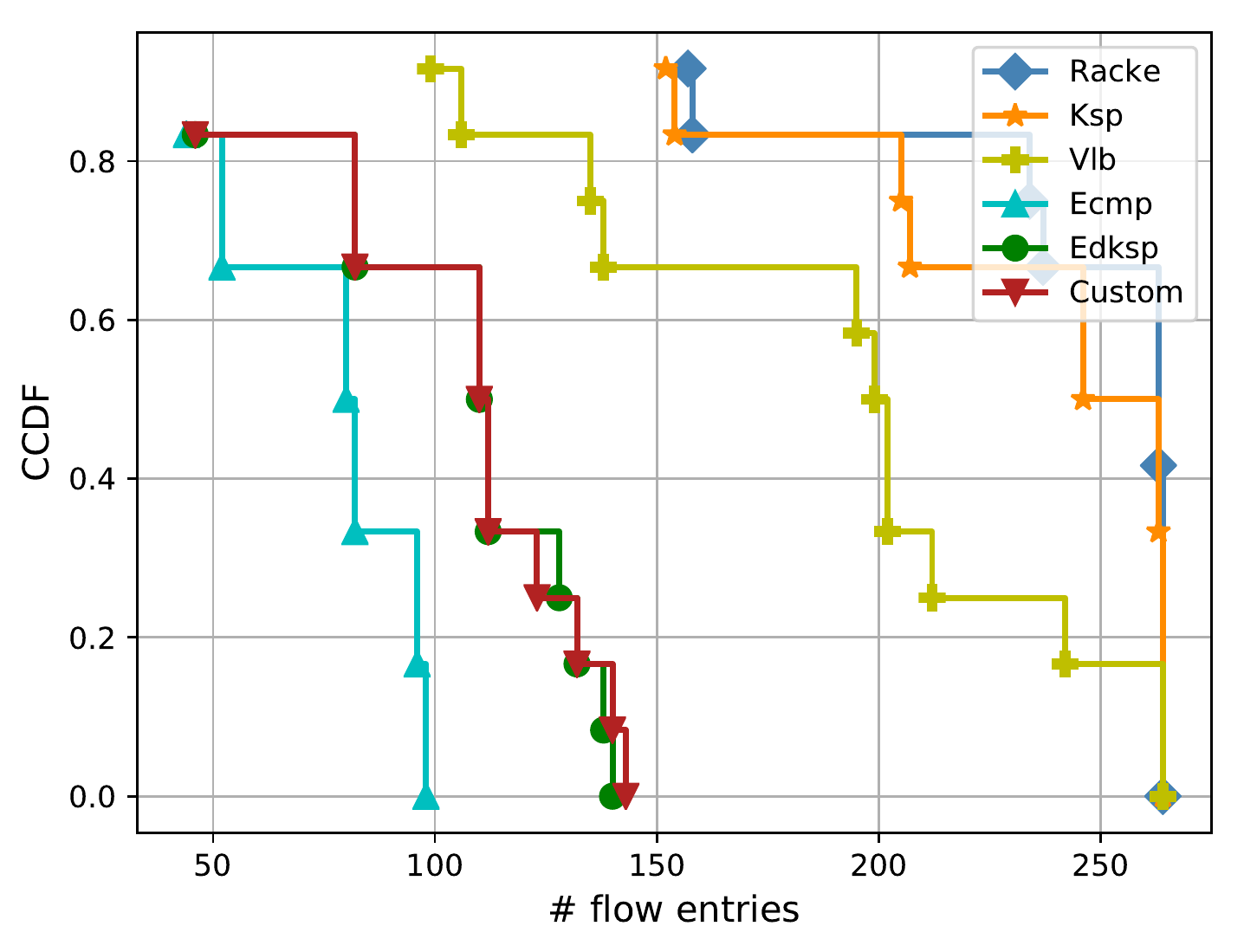}
            \caption[]
            {{\small program (G-Scale)}}
            \label{fig:flow-gscale-1-program-ccdf}
        \end{subfigure}
        \caption[ The average and standard deviation of critical parameters ]
        {\small \textbf{Algorithms vs. \# flow entries}, CCDF ($y=P(X>x)$) of number of flow entries in each router for each path computation method (``hardnop" is the method to select the maximum path budget that flow entries can bear; ``program" is to make full use of the flow
        entries).}
        \label{fig:algvsflowentry}
    \end{figure*}
\begin{figure*}[!htbp]
        \centering
        \begin{subfigure}[b]{0.325\textwidth}
            \centering
              \includegraphics[width=\linewidth]{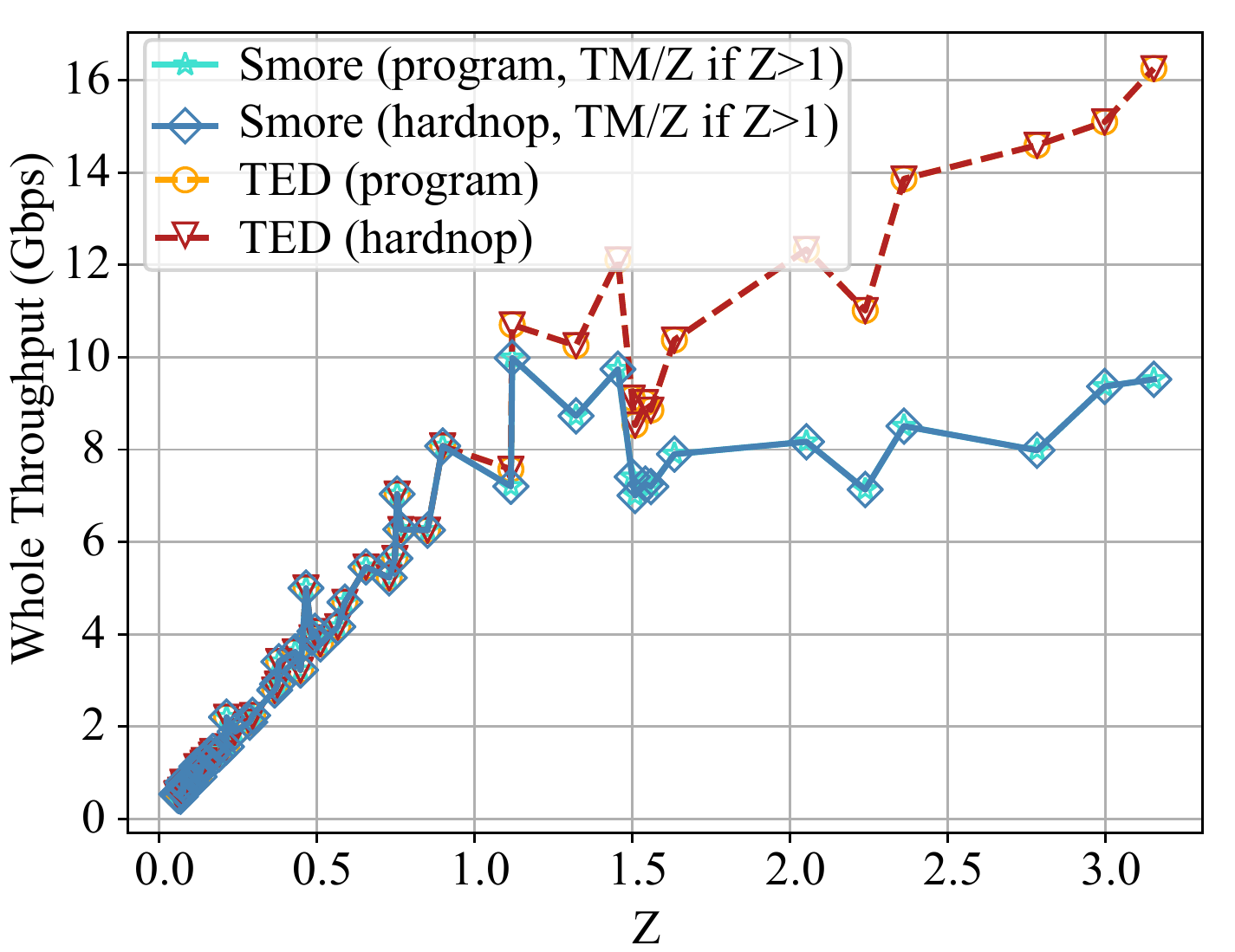}
            \caption[Network2]
            {{\small \textbf{TED vs. Smore \& hardnop vs. program} }}
                   \label{fig:gscale-Raeke-program-hardnop}
        \end{subfigure}
        \hfill
        \begin{subfigure}[b]{0.325\textwidth}
            \centering
            \includegraphics[width=\linewidth]{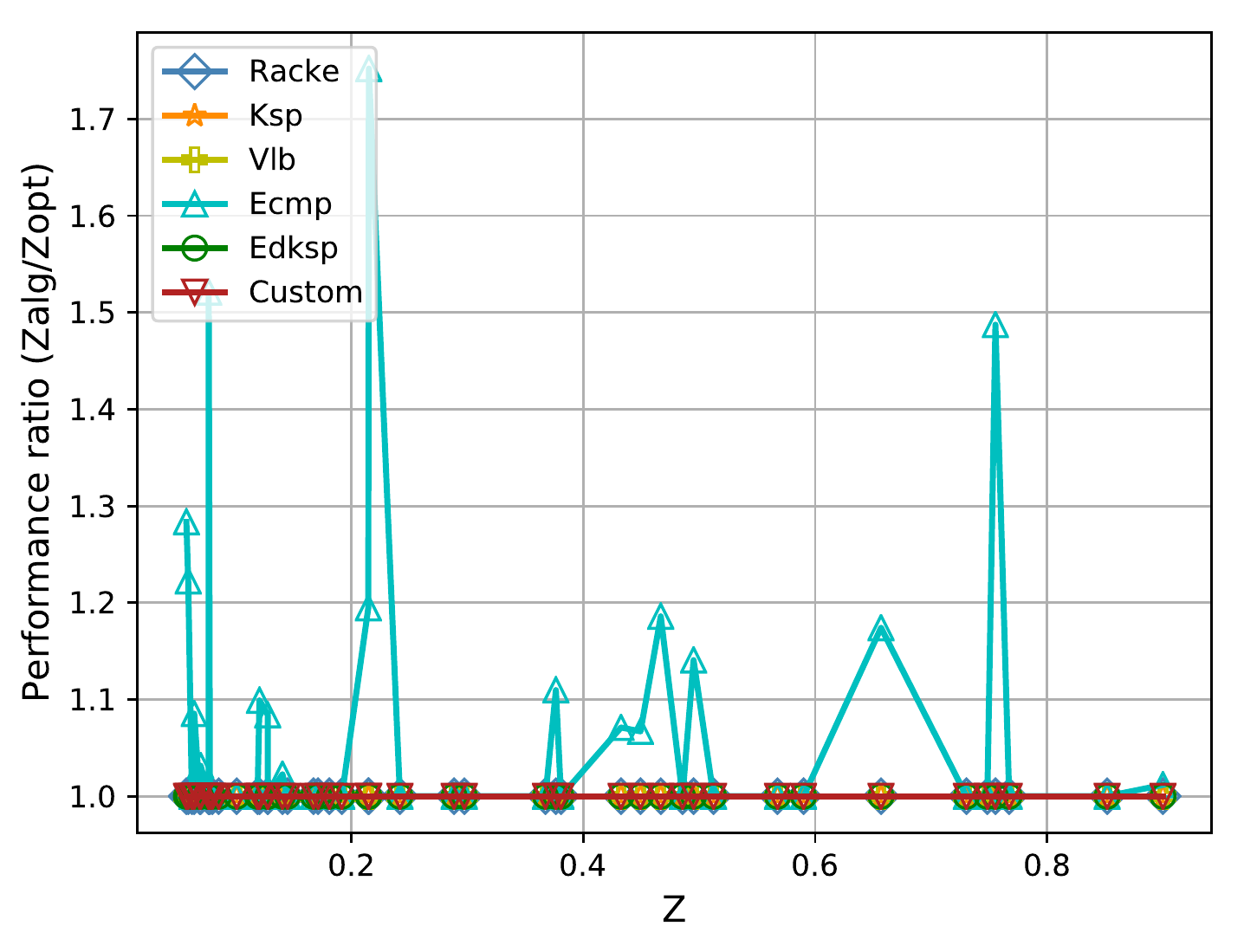}
            \caption[]
            {{\small \textbf{Performance ratio vs. algorithms}}}
            \label{fig:Z-gscale-1-hardnop-Zless1all}
        \end{subfigure}
        \hfill
        \begin{subfigure}[b]{0.325\textwidth}
            \centering
            \includegraphics[width=\linewidth]{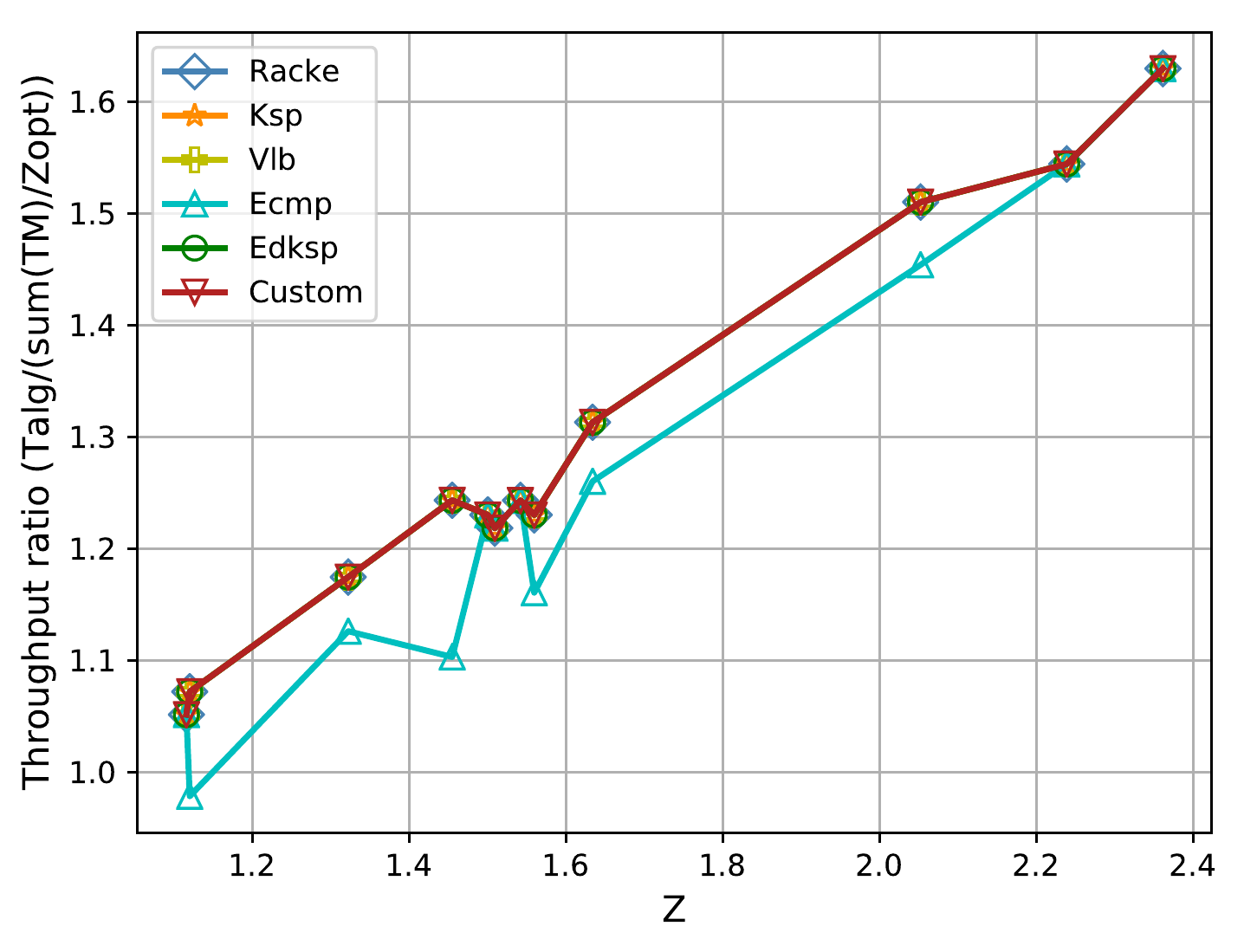}
            \caption[]
            {{\small \textbf{Throughput ratio vs. algorithms}}}
            \label{fig:Z-gscale-1-hardnop-Zlarge1-TZall}
        \end{subfigure}
        \caption[ The average and standard deviation of critical parameters ]
        {\small (a) Comparison of bandwidth allocation between TED and Smore using two path selection methods under the same flow entry limitation. (b) $Zopt \leq 1$, performance ratio and (c) $Zopt >1$, throughput ratio for each path computing method. (G-Scale)}
        \label{fig:performance}

    \end{figure*}

\begin{figure*}[!htbp]
        \centering
        \begin{subfigure}[b]{0.325\textwidth}
            \centering
            \includegraphics[width=\linewidth]{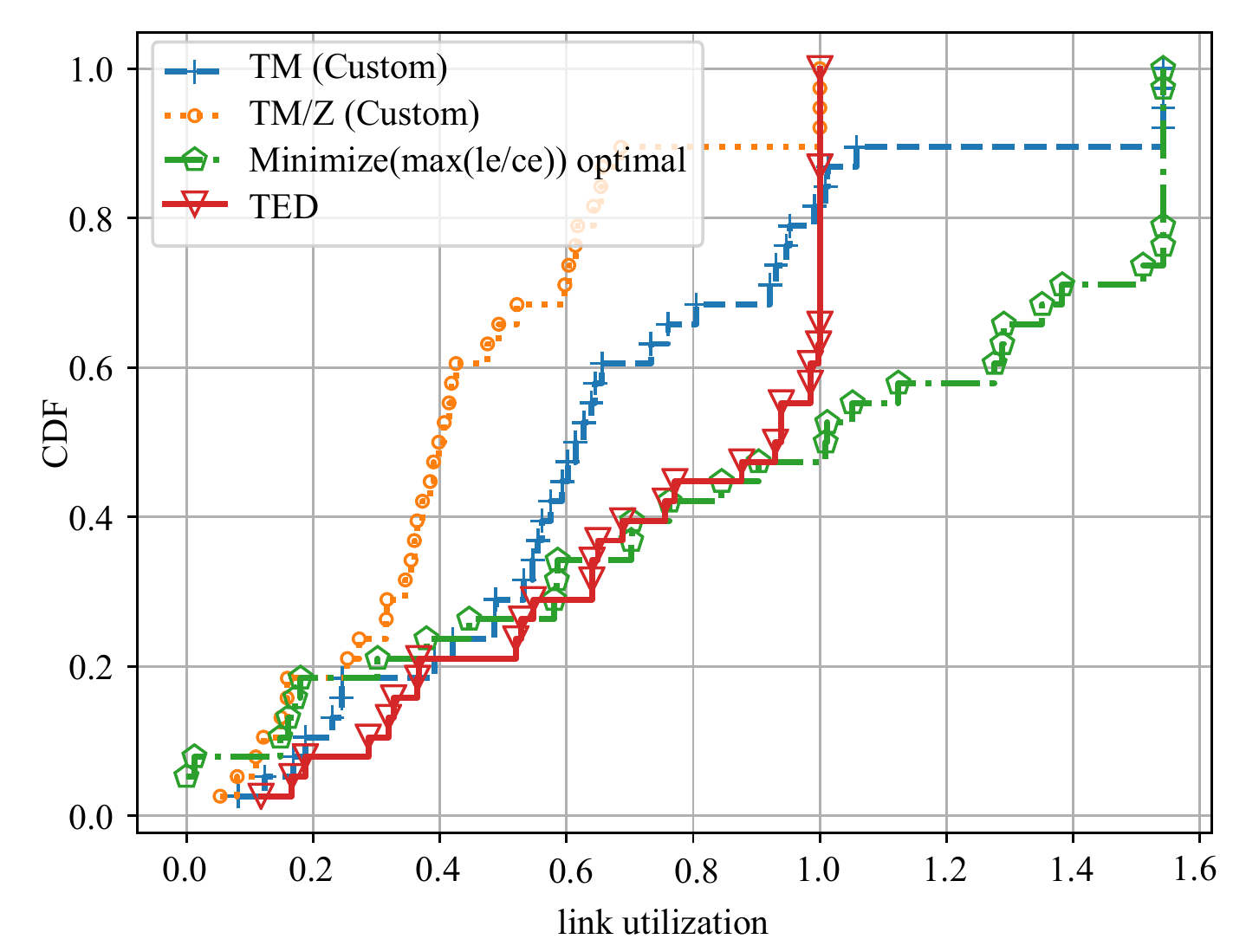}
            \caption[Network2]
            {{\small \textbf{TEs vs. link utilization}}}
            \label{fig:le-gscale-1-Custom-hardnop-cdfcomp1algTMZstep1-d48}
        \end{subfigure}
        \hfill
        \begin{subfigure}[b]{0.325\textwidth}
            \centering
            \includegraphics[width=\linewidth]{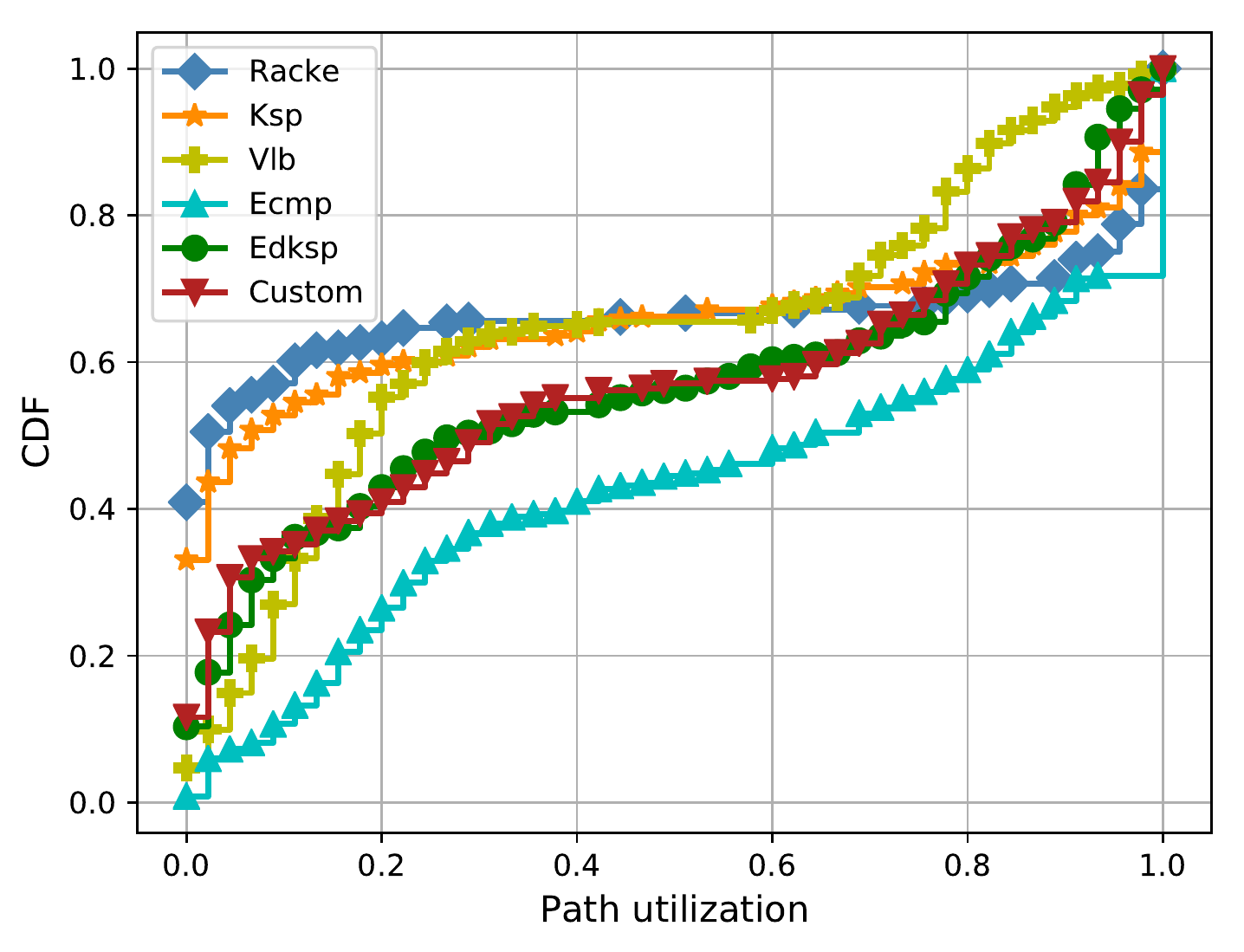}
            \caption[]
            {{\small \textbf{Path utilization vs. algorithms}}}
            \label{fig:Path-gscale-1-Zless1-hardnop-cdf}
        \end{subfigure}
        \hfill
        \begin{subfigure}[b]{0.325\textwidth}
            \centering
            \includegraphics[width=\linewidth]{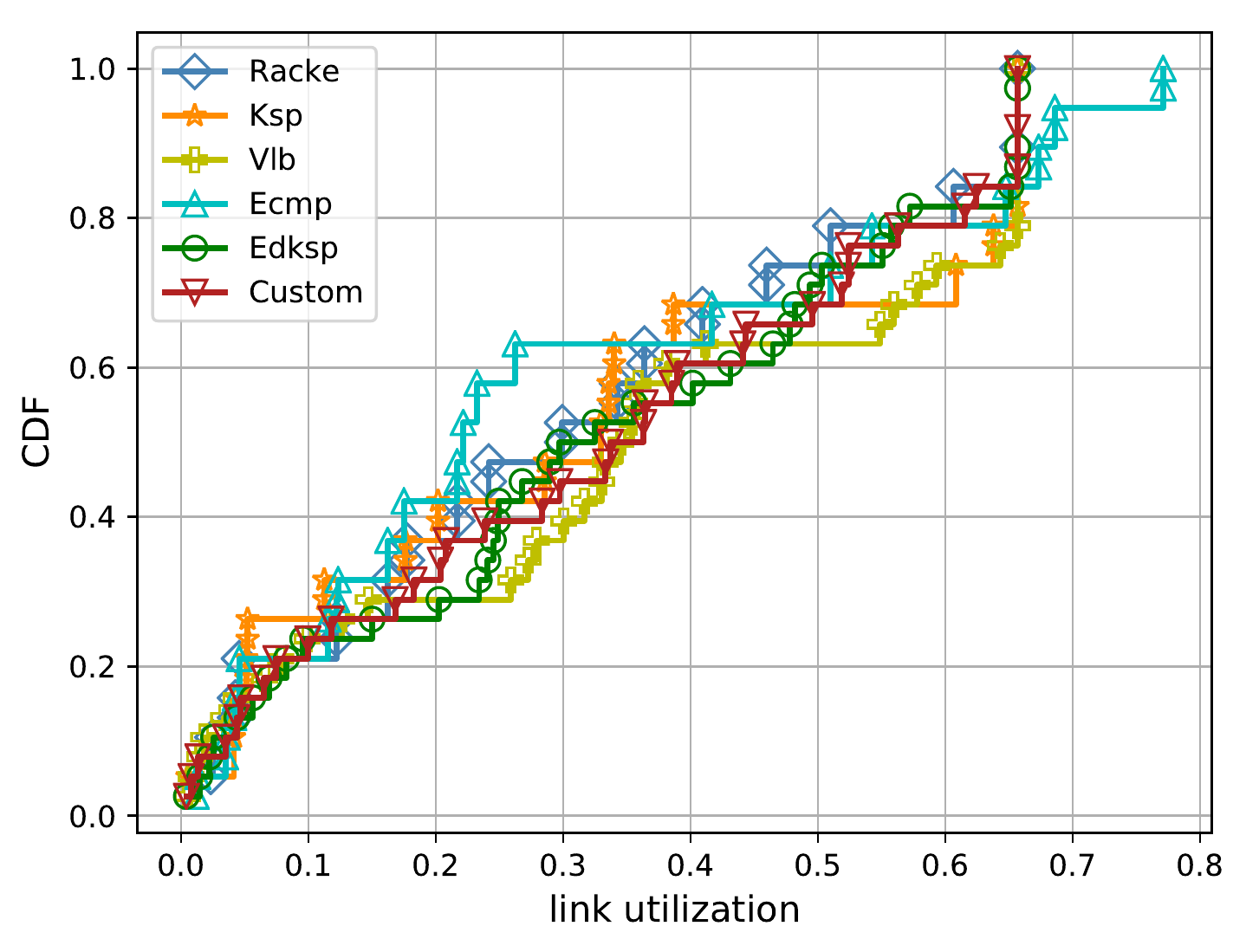}
            \caption[]
            {{\small \textbf{Link utilization vs. algorithms}}}
            \label{fig:le-gscale-1-hardnop-cdfalld41}
        \end{subfigure}
        \caption[ The average and standard deviation of critical parameters ]
        {\small (a) CDF ($y=P(X \leq x)$) of link utilization for TED using our maximum number of edge disjoint paths ($Zopt \approx 1.5$). (b) path utilization for $TM$s which make $Zopt \leq 1$,  and (c) $Zopt \approx 0.66$, link utilization for each path computing method. (G-Scale ``hardnop")}
        \label{fig:performance}
    \end{figure*}

\begin{figure*}
        \centering
        \begin{subfigure}[b]{0.325\textwidth}
            \centering
            \includegraphics[width=\linewidth]{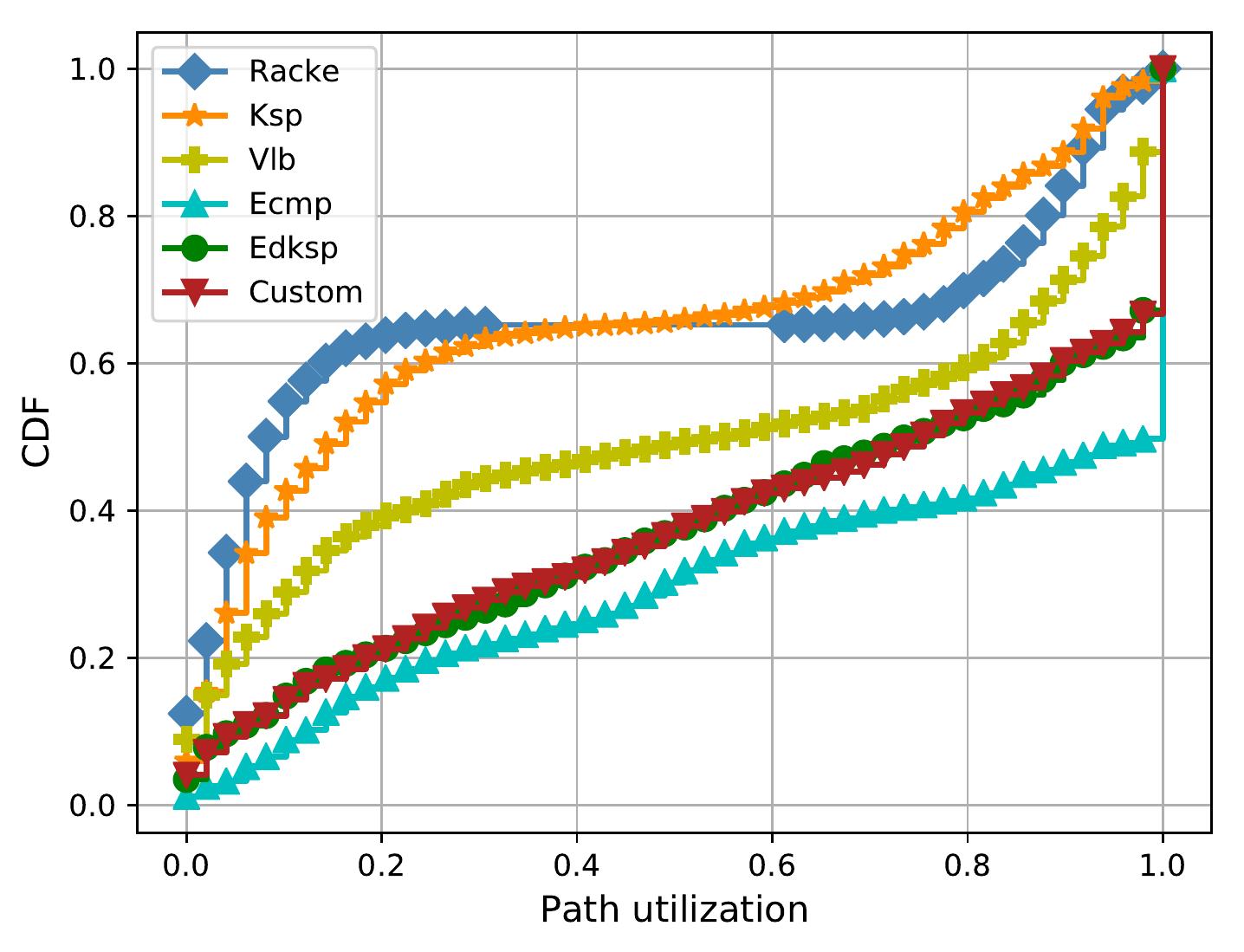}
            \caption[Network2]
            {{\small \textbf{Path utilization vs. algorithms}}}
            \label{fig:Path-Cernet-1-Zless1-hardnop-cdf}
        \end{subfigure}
        \hfill
        \begin{subfigure}[b]{0.325\textwidth}
            \centering
            \includegraphics[width=\linewidth]{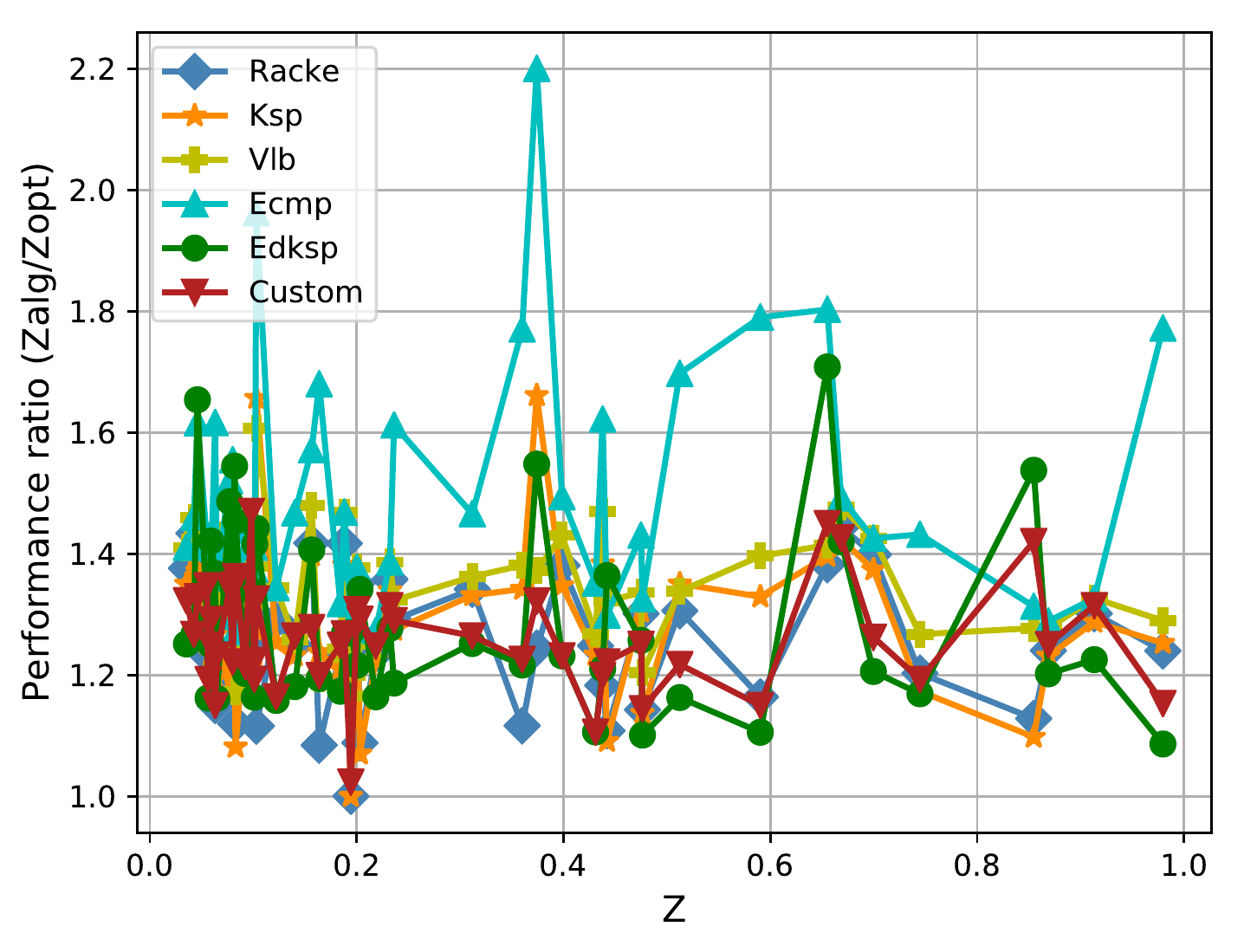}
            \caption[]
            {{\small \textbf{Performance ratio vs. algs
            (hardnop)}}}
            \label{fig:Z-Cernet-1-hardnop-Zless1all}
        \end{subfigure}
        \hfill
        \begin{subfigure}[b]{0.325\textwidth}
            \centering
            \includegraphics[width=\linewidth]{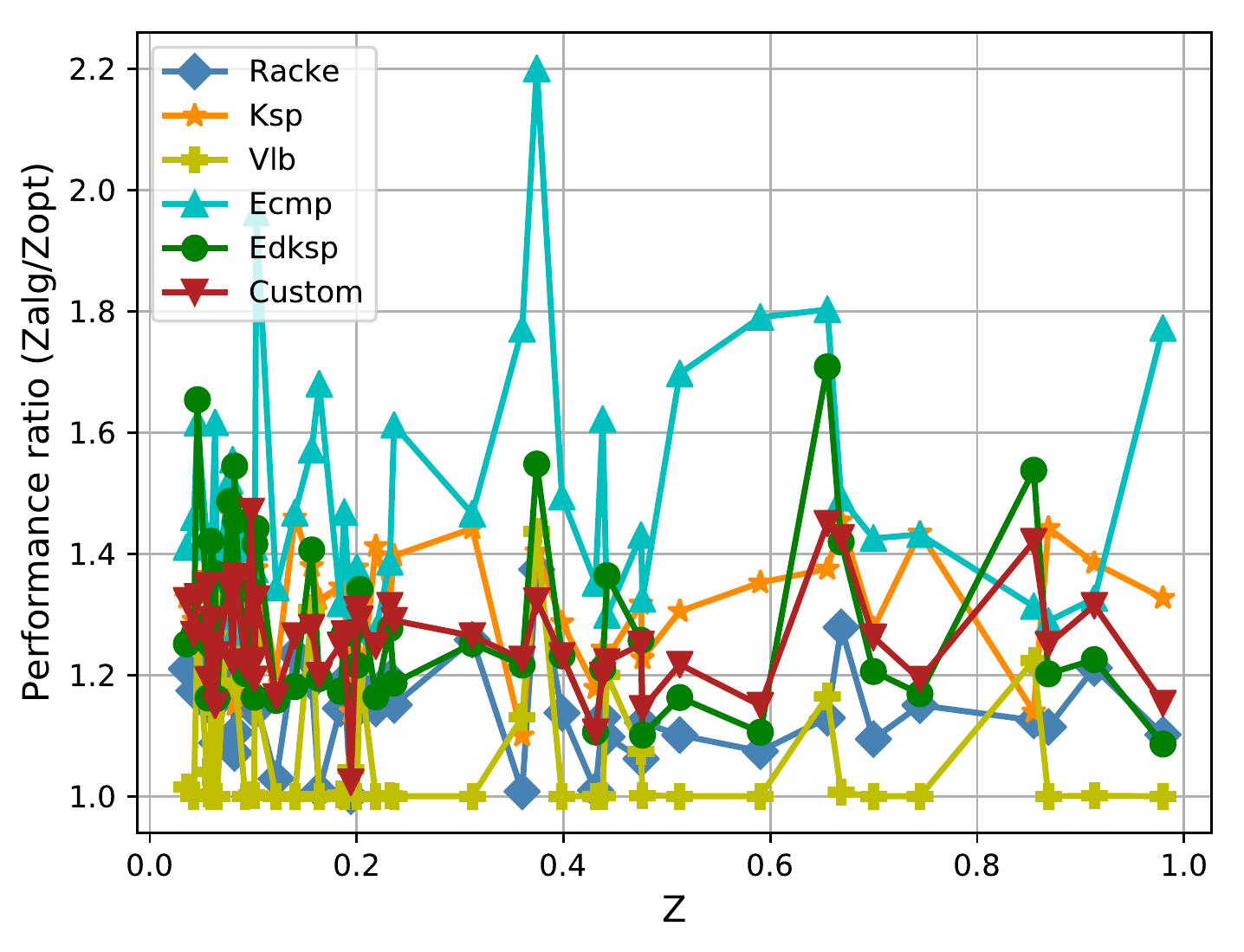}
            \caption[]
            {{\small \textbf{Performance ratio vs. algs (program)}}}
            \label{fig:Z-Cernet-1-program-Zless1all}
        \end{subfigure}
        \caption[ The average and standard deviation of critical parameters ]
        {\small (a) $Z \leq 1$, path utilization for each path computing method in Cernet (hardnop). (b) $Zopt \leq 1$, performance ratio for each path computing method in Cernet (hardnop). (c) $Zopt \leq 1$, performance ratio for each path computing method in Cernet (program).}
        \label{fig:performance}
    \end{figure*}

\begin{figure*}[!htbp]
\begin{subfigure}[b]{0.325\textwidth}
    \centering
\includegraphics[width=\linewidth]{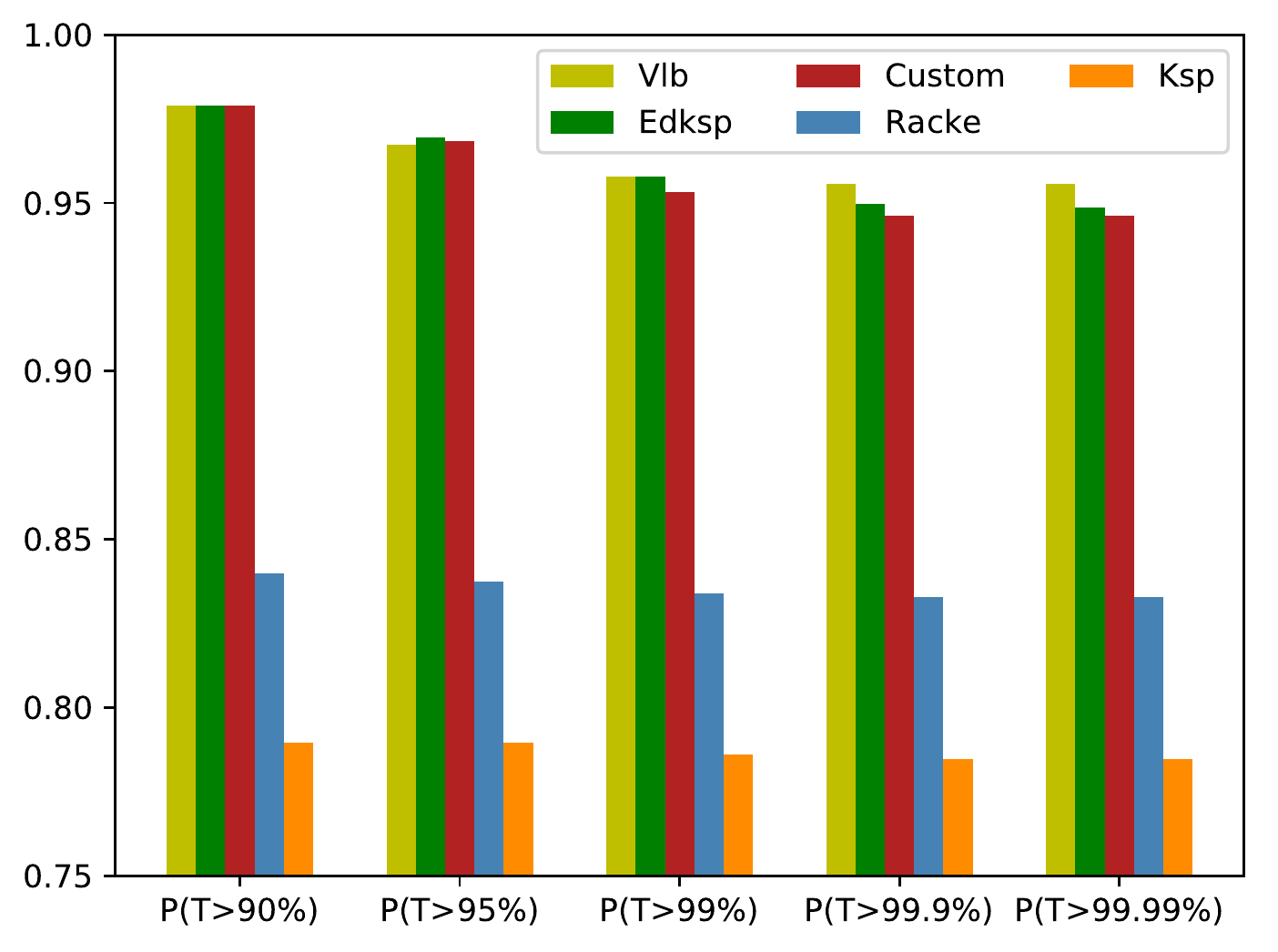}
\caption[]{\textbf{Robustness vs. algorithms}
}
\label{fig:TradiosinglelfailZless1}
\end{subfigure}
\hfill
\begin{subfigure}[b]{0.325\textwidth}
    \centering
\includegraphics[width=\linewidth]{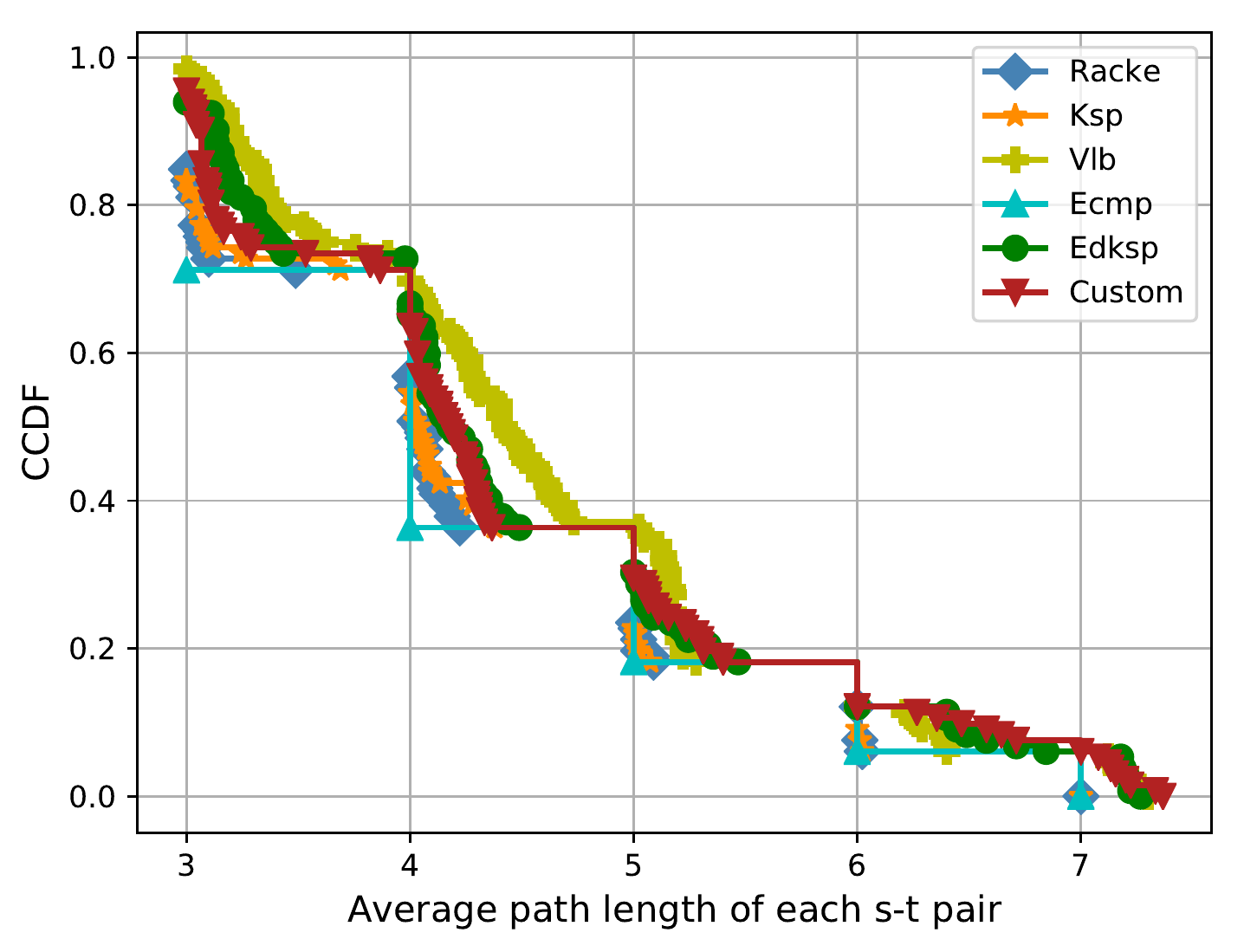}
\caption[]{\textbf{Path length vs. algorithms}
}
\label{fig:Pathlength-gscale-1-Zless1-hardnop1-cdf}
\end{subfigure}
\hfill
\begin{subfigure}[b]{0.325\textwidth}
    \centering
\includegraphics[width=\linewidth]{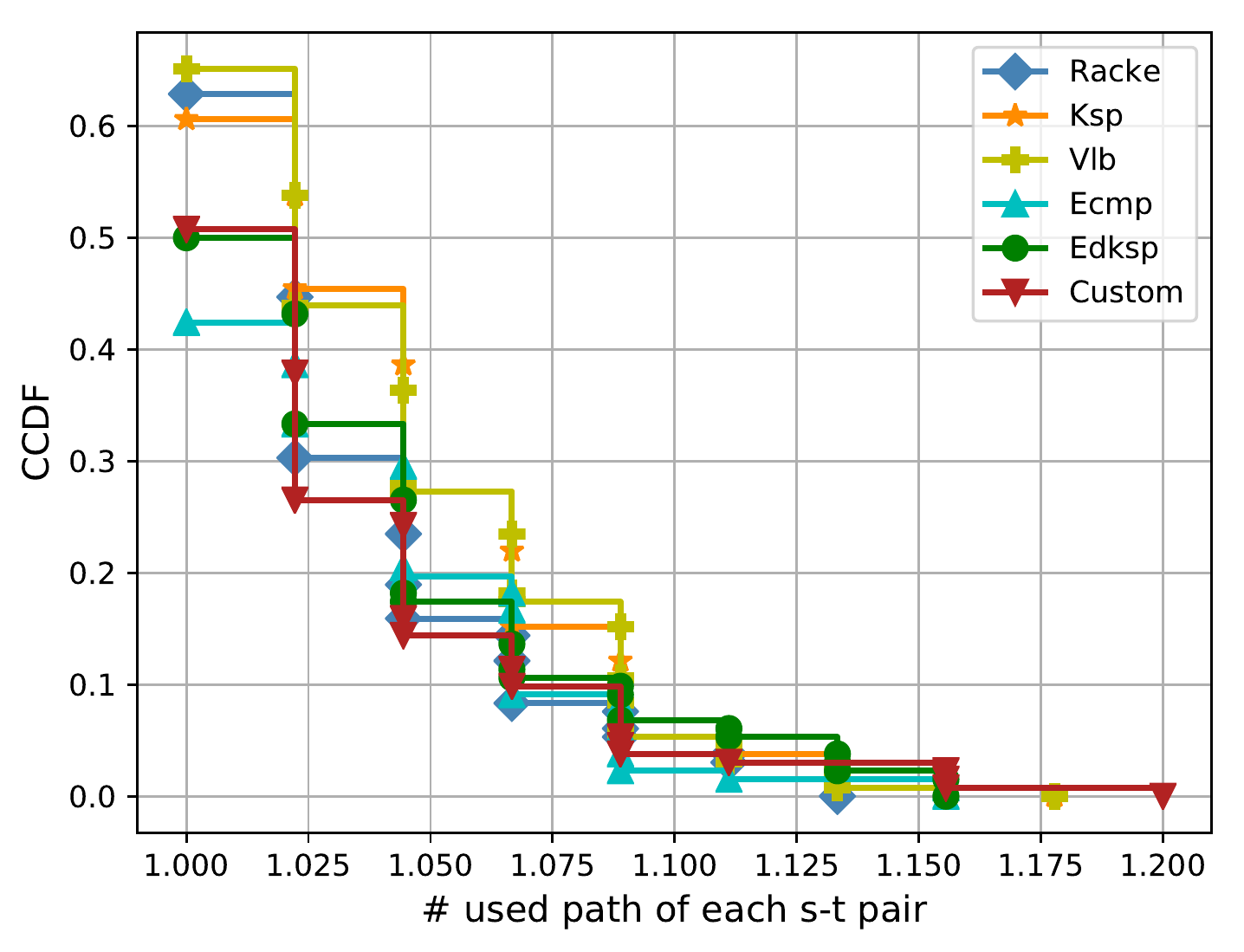}
\caption[]{\textbf{\# of used paths vs. algorithms}}
\label{fig:Pathnum-gscale-1-Zless1new-hardnop1-cdf}
\end{subfigure}
\caption[]{(a)$Zopt \leq 1$, throughput ratio after single edge failure for G-Scale ``hardnop"  (Racke stands for Smore). (b) $Zopt \leq 1$, the CCDF of average length of used paths (which have allocated bandwidth) for each path computing method in G-Scale (hardnop). (c) $Zopt \leq 1$, the CCDF of number of used paths between each $s-t$ pair for each path computing method in G-Scale (hardnop).}
    \end{figure*}

\subsection{Setup}

\emph{Methodology.} We vary demands, single link failures and topologies to compare path computation algorithms, path selection algorithms and TE optimizations.

\emph{TM Generation.} We use gravity model \cite{ref:genTM} to generate 60 traffic matrices randomly.

\emph{Topologies.}
We select 20 topologies from topology zoo \cite{ref:topologyzoo}, which are also used in Yates \cite{ref:yates}.

\emph{Path set computation algorithms.}
Our algorithm based on Dinic to compute the maximum number of edge disjoint paths is called ``Custom" in the figures (implemented in C++). We compare with Racke, Ksp, Vlb, Ecmp, Edksp which are implemented in OCaml by Yates\cite{ref:yates}. Racke stands for Racke's oblivious routing algorithm used in Smore \cite{ref:smore}. Ksp stands for Yen's algorithm to compute k-shortest paths which is commonly used in TE. Vlb \cite{ref:vlb} means Valiant Load Balancing which routes traffic via randomly selected intermediate nodes. Ecmp is widely used equal-cost multi-path routing. Edksp is short for edge-disjoint $k$-shortest paths.

\emph{Path selection algorithms.}
Our path selection algorithm is denoted as ``hardnop" in the figures (select the maximum path budget that flow entries can hold). We compare it with another method designed by us, called ``program". It is a two-step path selection method to find path selection result by making full use of all flow entries. We implement both of them in Julia, and apply the same flow entries' limitation to both methods.

\emph{TE implementation.} We compare TED with optimal MCF \cite{ref:optimalmcf}, TE (TM) and TE (TM/Z, when $Z>1$). Their target is to minimize the maximum link utilization. Optimal MCF does not have path or flow entry limitation (does not need to input paths as it can use the whole network under the flow conservation constraint). TE (TM) is the unmodified TE with objective of minimizing the maximum link utilization used in Smore \cite{ref:smore}. TE (TM/Z, when $Z>1$) is the simple modified TE, when $Z>1$, to use $TM/Z$ as the bandwidth allocation result to guarantee no congestion. We implement them in Julia by calling Gurobi's optimization solver.

We use the following metrics for performance evaluation. When $Zopt \leq 1$ ($Zopt$ means the maximum link utilization result of optimal MCF), we evaluate the performance ratio ($= Zalg/Zopt$) \cite{ref:perfratio}; and when $Zopt>1$, we use the throughput ratio ($= Talg/(sum(TM)/Zopt)$). $Zalg$ and $Talg$ is computed using TED's architecture and using corresponding ``alg" as the algorithm to compute the path set in Phase I. Performance ratio show that how far from the $Zalg$ to $Zopt$.  Throughput ratio shows that how much improve of $Talg$ compared with $sum(TM)/Zopt$ (simple method to use $TM/Zopt$ guaranteeing no congestion when $Zopt>1$). The source code of our experiment is on \textcolor{blue}{https://github.com/ylxdzsw/TEexp}.

\subsection{Number of flow entries}

We use much less number of flow entries (Fig. \ref{fig:flow-gscale-1-hardnop-ccdf}) than Vlb, Ksp and Racke, but more than Ecmp for Gscale. Later we will show our robustness and performance is competitive with using other algorithms and much better than Ecmp. This is important for real networks which may not have that much flow entries. Especially to notice that large networks to guarantee good all-to-all communication need more than twice flow entries for Ksp and Racke compared with our Custom (Fig. \ref{fig:flow-cernet-1-hardnop-ccdf}).

Also, comparing with two-step path selection method, ``program" (Fig. \ref{fig:flow-gscale-1-program-ccdf}), Racke, Ksp and Vlb use more flow entries than our path selection method, ``hardnop" (Fig. \ref{fig:flow-gscale-1-hardnop-ccdf}). This is because ``program" can make full use of the flow entries and for Racke, Ksp and Vlb, their path sets have more paths to select than Ecmp, Edksp and Custom. However, next we will show that ``program" does not have better performance than ``hardnop" for each path set computing algorithms at least for G-Scale-like topologies.

\subsection{Efficiency}
\begin{figure}[!htbp]
  \centering
\includegraphics[width=0.7\linewidth]{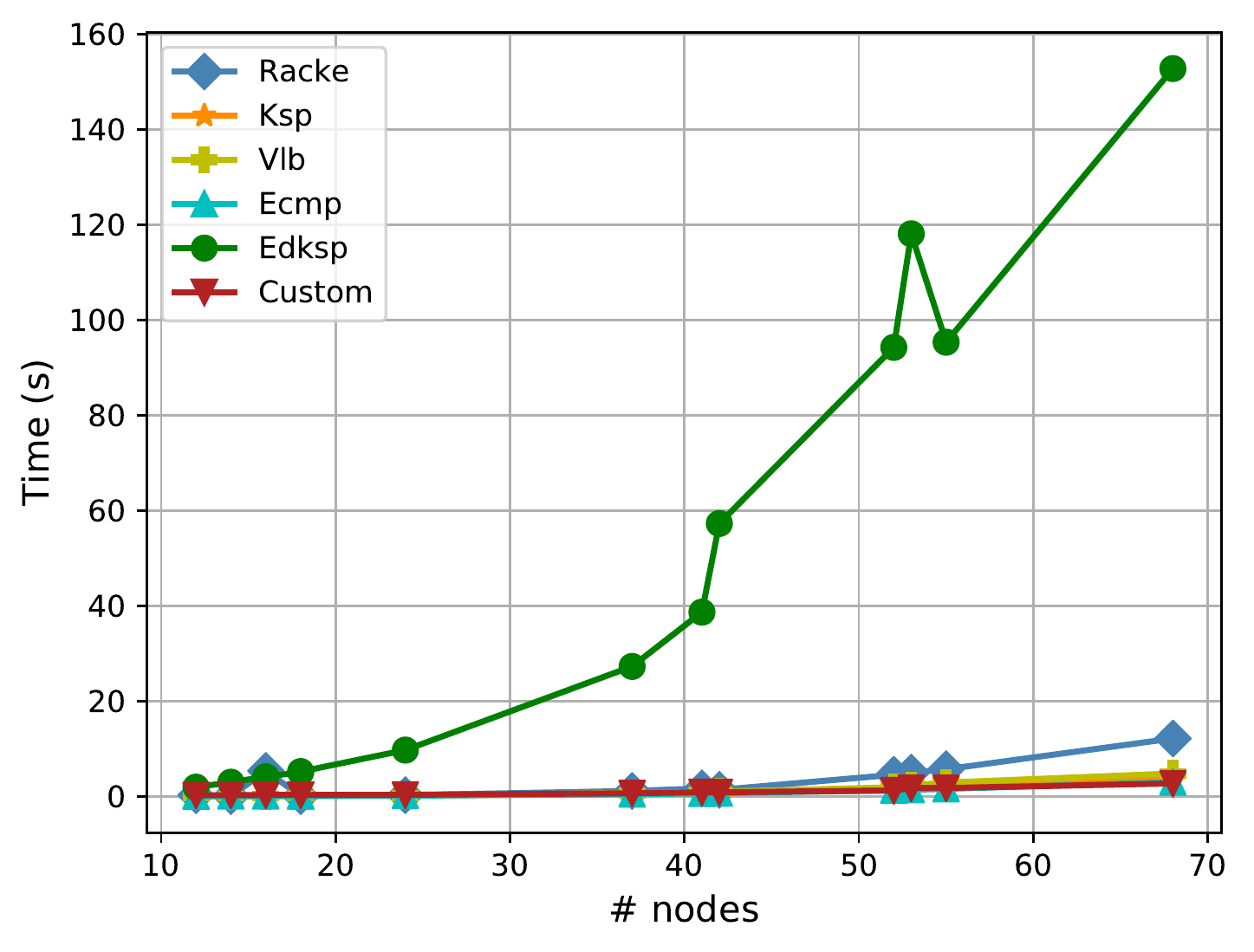}
\caption[]{\textbf{Computation time vs. the scale of the topology}}
\label{fig:algtime}
\end{figure}

Fig. \ref{fig:algtime} shows the computation of various algorithms. We can see that our path computation algorithm Custom is faster than Racke and much faster than Edksp.  It is especially important when failure happens, to compute new paths in order to guarantee packet loss only  lasts for a short time.

\subsection{Performance}

Our TED's overall throughput is same as Smore (Racke) when $Z \leq 1$. We also consider $Z>1$ in our Phase IV of Fig. \ref{fig:sysnew4} which is not considered in Smore. Therefore, we set the bandwidth allocation to be $TM/Z$ when $Z>1$ for Smore in order to avoid congestion. As TED can make full use of the link capacity when $Z>1$ (Fig. \ref{fig:le-gscale-1-Custom-hardnop-cdfcomp1algTMZstep1-d48}), it is better than Smore (TM/Z when $Z>1$). Also, Smore and TED using ``program" as the path selection method do not show improvement in G-Scale compared with our TED's path selection method, ``hardnop". That is also the reason that we use ``hardnop" instead of ``program" for G-Scale like well connected topologies. The other reason is that ``program" needs to solve two 0-1 integer linear programming (NP-complete problem).

No matter $Zopt \leq 1$ (Fig. \ref{fig:Z-gscale-1-hardnop-Zless1all}) or $Zopt>1$ (Fig. \ref{fig:Z-gscale-1-hardnop-Zlarge1-TZall}), performance of TED (Custom) is much better than Ecmp and competitive with other path set computing algorithms (all algorithms use TED's architecture) while using less flow entries. The reason is that TED has higher path utilization shown in Fig. \ref{fig:Path-gscale-1-Zless1-hardnop-cdf}. Note that if we remove the 40\% almost unused paths for Racke and Ksp, although their used flow entries can be less, their performance after failures can be even worse. Also, Fig. \ref{fig:le-gscale-1-hardnop-cdfalld41} shows that when $Zopt \approx 0.66$, the reason of Fig. \ref{fig:Z-gscale-1-hardnop-Zless1all}'s peak ($Zecmp/Zopt$ is about 1.18) of Ecmp is  $Zecmp \approx 0.78$ ($1.18=0.78/0.66$). Notice that about 20\% Ecmp links have higher link utilization than $Zopt$ which is because Ecmp has less paths to balance the traffic.

\subsection{Robustness}

To test robustness, we fail each link once. For each link failure, we re-run TED using the changed topology (includes all the Phases) and the path set is computed by each algorithm.
Although we use less flow entries, our performance after failure is better than Racke, Ksp and much better than Ecmp (which is zero in Fig. \ref{fig:TradiosinglelfailZless1}, so we remove it) when $Zopt \leq 1$. It is because we use edge-disjoint paths so that any single link failure at most makes one path unavailable for the influenced $s-t$ pairs. We are competitive with Edksp which uses edge-disjoint $k$-shortest paths and Vlb which routes traffic via randomly selected intermediate nodes but remember that we are much faster than Edksp and use less flow entries than Vlb.

\subsection{Path length}

As shown in Fig. \ref{fig:Pathlength-gscale-1-Zless1-hardnop1-cdf}, the average length of used paths between each $s-t$ pair of TED (Custom) is only a slightly longer than Ecmp, Ksp and Racke. Custom, Edksp and Vlb only make less than 10\% $s-t$ pairs has longer (less than 0.5 hop) average used path length than the longest path of Racke, Ecmp and Ksp for G-Scale.

\subsection{Limitations and future works}
One shortcoming of our edge disjoint paths is that Custom may not perform very well when some $s-t$ pairs in a topology have only one edge disjoint path. We believe that this situation is not common, because we found that the 40 topologies in topology zoo we evaluated have rich connections.

The other limitation is that for network with diverse link capacity, using only edge disjoint paths may not full utilize the capacity. In the future, we will explore how to select robust paths under such condition.

We would also point out some interesting results which need further  investigation.
For example, Fig. \ref{fig:Pathnum-gscale-1-Zless1new-hardnop1-cdf} shows that for all the path computation methods, at least more than 30\% $s-t$ pairs have only one path with allocated bandwidth, although on average each $s-t$ pair has three paths inputted to the TE optimization in G-Scale. We will attempt to analyze the reason theoretically. Fig. \ref{fig:Path-Cernet-1-Zless1-hardnop-cdf} shows that about 30\% paths are used for all $TM$s for Custom and Edksp in Cernet but our performance ratio is still close to Racke using hardnop (Fig. \ref{fig:Z-Cernet-1-hardnop-Zless1all}). One possible reason is that edge-disjoint paths can use less paths to cover more important links than Racke. In this case, we need to find and protect such important links specially. The other interesting finding is that combined with our program, Vlb performs near optimal for Cernet (Fig. \ref{fig:Z-Cernet-1-program-Zless1all}) except the paths can be longer.

\section{Related Work}
\label{sec:related}

\noindent \textbf{Path selection for multi-commodity flow:}
Multi-commodity flow problem \cite{ref:networkflow} has been studied for many years. Recently, Merlin \cite{ref:merlin} and SNAP \cite{ref:snap} use a mixed-integer linear program to select one path for each ingress-egress switch pair which is not enough for WANs. Danna et al. \cite{ref:fairness} uses ``water filling'' related algorithm to allocate bandwidth for multi-commodity flow. It is very efficient and stable to the variance of demand. However, as it focuses on fairness, in their setting, each commodity has multiple possible paths to route its demand, how to select these paths is not explained. Caesar et al. \cite{ref:dyn} argues that decoupling failure recovery from path computation leads to networks that are inherently more efficient, more scalable and easier to manage. We were inspired by their work although they only propose a multi-path scheme that endpoint utilizes a fixed set of $k$ disjoint-as-possible available paths without mentioning how to select these paths, either.

\noindent \textbf{Improve reliability for data center WAN:}
The most widely used approach to deal with network failures, including link or switch failures, is to re-compute a new TE solution based on the changed topology and re-program the switches \cite{ref:b4,ref:swan}. However, re-computing a new TE plan and updating the forwarding rules across the entire network take at least minutes and are error-prone.

Several proactive approaches have been proposed to solve this important problem. Suchara et al. \cite{ref:joint} modifies the rescaling behavior of ingress switch by pre-computing and configuring forwarding rules based on the likelihood of different failure cases to prevent rescaling-induced congestion after a data plane fault. Although it achieves near-optimal load balancing, this approach can handle only a limited number of potential failure cases as there are exponential many of them to consider. SWAN \cite{ref:swan} develops a new technique that leverages a small amount of scratch capacity on links to apply updates in a provably congestion-free manner. FFC \cite{ref:ffc} is proposed to proactively protect a network from congestion and packet loss due to data and control plane faults. Although FFC spreads network traffic such that congestion-free property is guaranteed under arbitrary combinations of up to $k$ failures, the price is very high. About 5\%-10\% of the network capacity depending on $k$ has to be always left vacant to handle traffic from rescaling.

All of them either uses the k-shortest paths for each ingress-egress switch pair, or selects  paths considering the failure probability. The SWAN's \cite{ref:swan} allocation function allocates rate by invoking TE separately for classes in priority order. After a class is allocated, its allocation is removed from remaining link capacity. Doing like this, SWAN also ensures that higher priority traffic is likelier to use shorter paths. However, it still has not mentioned how to select better paths as the input of TE. Then Smore \cite{ref:smore} is proposed to use semi-oblivious traffic engineering. It works well for non-peak hours when flow tables are enough but works worse than TED when flow table limit is the same as TED. Smore does not consider how to set the suitable path budget and how to spread traffic when $TM$ is large. As we have shown in evaluation, Smore's  performance is worse than TED for heavily-loaded networks before and after failures and it takes longer time to compute the paths.

\section{Conclusion}
\label{sec:conclusion}
{}

This paper presents the motivation, design, and evaluation of TED, a scalable TE system for SDN. We present our four-phase approach to guarantee network performance and robustness, no matter how large the traffic matrix is, under the limitation of flow entries. TED uses the maximum number of edge disjoint paths between each $s-t$ pair as the path set. TED then select paths from the path set by computing the maximum path budget to guarantee paths are diverse enough to handle various traffic matrices. Next, we use minimizing maximum link utilization optimization to compute $Z$. If $Z \leq 1$, set the weights of paths. Otherwise, set the weights and allocated bandwidth re-computed by the maximizing whole throughput optimization under the $TM/Z \leq TM' \leq TM$ constraint (to guarantee each $s-t$ pair can at least get $d_{s,t}/Z$ bandwidth). Such process guarantees not to sacrifice performance or QoE to provide robustness. TED utilizes priority queue to guarantee the transmission of loss and delay sensitive traffic before and after rescaling \cite{ref:kuijia18}. TED is also fast to guarantee that once failure happens, after rescaling, it can re-compute the weights and bandwidth allocation immediately using the changed topology. Also, operators can change the flow entries' limitation or capacity limitation to some percent to trade-off between performance and robustness.

\label{ConcPage}

{\small
\bibliographystyle{abbrv}
\balance
\bibliography{main}
}

\end{document}